\documentclass[floatfix,amssymb,preprint,aps,pra, superscriptaddress,showpacs]{revtex4-2}
\usepackage{amsmath}
\usepackage{amsfonts}
\usepackage{amssymb}
\usepackage{graphicx}
\usepackage{units}
\usepackage{color}
\usepackage{xspace}
\usepackage[normalem]{ulem}
\usepackage{textcomp}

\DeclareUnicodeCharacter{2009}{\,}
\DeclareUnicodeCharacter{2212}{-}

\begin{document}

\title{Tailored 1D/2D Van der Waals Heterostructures for Unified Analog and Digital Electronics}

\setcounter{footnote}{1}
\author{Bipul Karmakar}
\altaffiliation{These authors contributed equally to this work}
\affiliation{School of Physical Sciences, Indian Association for the Cultivation of Science, 2A $\And$ B Raja S. C. Mullick Road, Jadavpur,
Kolkata 700032, India}
\author{Bikash Das}
\altaffiliation{These authors contributed equally to this work}
\affiliation{School of Physical Sciences, Indian Association for the Cultivation of Science, 2A $\And$ B Raja S. C. Mullick Road, Jadavpur,
Kolkata 700032, India}
\author{Shibnath Mandal}
\affiliation{School of Physical Sciences, Indian Association for the Cultivation of Science, 2A $\And$ B Raja S. C. Mullick Road, Jadavpur,
Kolkata 700032, India}
\author{Rahul Paramanik}
\affiliation{School of Physical Sciences, Indian Association for the Cultivation of Science, 2A $\And$ B Raja S. C. Mullick Road, Jadavpur,
Kolkata 700032, India}
\author{Sujan Maity}
\affiliation{School of Physical Sciences, Indian Association for the Cultivation of Science, 2A $\And$ B Raja S. C. Mullick Road, Jadavpur,
Kolkata 700032, India}
\author{Tanima Kundu}
\affiliation{School of Physical Sciences, Indian Association for the Cultivation of Science, 2A $\And$ B Raja S. C. Mullick Road, Jadavpur,
Kolkata 700032, India}
\author{Soumik Das}
\affiliation{School of Physical Sciences, Indian Association for the Cultivation of Science, 2A $\And$ B Raja S. C. Mullick Road, Jadavpur,
Kolkata 700032, India}
\author{Mainak Palit}
\affiliation{School of Physical Sciences, Indian Association for the Cultivation of Science, 2A $\And$ B Raja S. C. Mullick Road, Jadavpur,
Kolkata 700032, India}
\author{Koushik Dey}
\affiliation{School of Physical Sciences, Indian Association for the Cultivation of Science, 2A $\And$ B Raja S. C. Mullick Road, Jadavpur,
Kolkata 700032, India}
\author{Kapildeb Dolui}
\affiliation{Lomare Technolgies Limited, 6 London Street, London EC3R 7LP, United Kingdom}
\affiliation{Department of Materials Science $\And$ Metallurgy, University of Cambridge, 27 Charles Babbage Road, Cambridge CB3 0FS, United Kingdom}
\author{Subhadeep Datta}
\altaffiliation{sspsdd@iacs.res.in}
\affiliation{School of Physical Sciences, Indian Association for the Cultivation of Science, 2A $\And$ B Raja S. C. Mullick Road, Jadavpur,
Kolkata 700032, India}

\begin{abstract}
    \bfseries{ 
We report a sequential two-step vapor deposition process for growing mixed-dimensional van der Waals (vdW) materials, specifically Te nanowires (1D) and MoS$_2$ (2D), on a single SiO$_2$ wafer. Our growth technique offers a unique potential pathway to create large scale, high-quality, defect-free interfaces. The assembly of samples serves a twofold application: first, the as- prepared heterostructures (Te NW/MoS$_2$) provide insights into the atomically thin depletion region of a 1D/2D vdW diode, as revealed by electrical transport measurements and density functional theory-based quantum transport calculations. The charge transfer at the heterointerface is confirmed using Raman spectroscopy and Kelvin probe force microscopy (KPFM). We also observe modulation of the rectification ratio with varying applied gate voltage. Second, the non-hybrid regions on the substrate, consisting of the as-grown individual Te nanowires and MoS$_2$ microstructures, are utilized to fabricate separate p- and n-FETs, respectively. Furthermore, the ionic liquid gating helps to realize low-power CMOS inverter and all basic logic gate operations using a pair of n and p- field-effect transistors (FETs) on Si/SiO$_2$ platform.  This approach also demonstrates the potential for unifying diode and CMOS circuits on a single platform, opening opportunities for integrated analog and digital electronics.}
    \end{abstract}
\maketitle

\section{Introduction}
 
In very-large-scale integration (VLSI), microprocessors often require separate components for timing control, protection, interrupt handling, and clock generation, increasing both power consumption and chip area \cite{reis2006design}. Materials like silicon are used for CMOS-based timing circuits, silicon-on-insulator (SOI) and copper interconnects for clock-pulse operation and signal routing, and silicon carbide (SiC) for over-voltage protection, but these approaches are less efficient as device scales shrink \cite{Cao2023}. Embedded systems with quartz-based real-time clocks (RTCs) further complicate microprocessor design. Following the achievement of sub-10 nm channel thresholds in silicon logic technologies \cite{akinwande2019graphene}, 3D integration of integrated circuits (ICs) has become a primary goal to overcome the limitations of traditional 2D scaling in the `More Moore' era \cite{pendurthi2024monolithic,jayachandran2024three}. Researchers are particularly focused on integrating opposite polarity field effect transistors (FETs), junctions, and interconnects on a single platform, which enable simultaneous operation of mixed (analog and digital) signals. In the realm of non-silicon materials, recent demonstrations of 3D stacking with van der Waals (vdW) MoS$_2$, MoTe$_2$, WSe$_2$, etc. mark significant progress in the 3D integration of complementary FETs \cite{xia2022wafer,liu2023large,pendurthi2024monolithic,jayachandran2024three}. By choosing a 1D/2D heterostructure template instead of a 2D/2D configuration, one can gain the advantage of constructing interconnects and FETs using the 1D material for IC operations. Additionally, transport phenomena occurring at the atomic scale of 1D/2D junctions continue to be a key area of interest for fundamental researchers. For example, the effective depletion region in an atomically thin p-n junction remains a subject of debate.
  
vdW materials offer a promising avenue for exploring next generation devices, both for fundamental research, such as superconductivity \cite{navarro2016enhanced} and practical applications such as FETs and logic devices  \cite{radisavljevic2011single,jariwala2014emerging}. Their weak interlayer bonding between the adjacent layers creates a defect-free interface, circumventing the lattice matching constraints typically encountered in epitaxial heterojunctions \cite{jariwala2017mixed, geim2013van,haigh2012cross}. Transition-metal dichalcogenides (TMDCs) are highly celebrated within the family of 2D materials (after graphene and hexagonal boron nitride (hBN)) for their atomically smooth surfaces, high mobility, tunable layer-dependent bandgaps, controlled layer-by-layer stacking, functionalization, strain engineering, and more \cite{kang2013band,mak2010atomically,splendiani2010emerging,radisavljevic2011single}. . 
To facilitate a range of analog and digital circuit operations, the most effective strategy is to arrange p-type devices, n-type devices, and heterojunctions between them in close proximity on a single substrate. However, the main challenge is identifying a suitable p-type alternative to construct a vdW p-n heterojunction, since most 2D TMDCs are inherently n-type due to their intrinsic anion vacancy sites. One potential solution is the use of 2D black phosphorus (BP) in various system for the fabrication of p-n junction \cite{miao2017vertically, liu2017modulation, srivastava2019van}. However, issues such as poor environmental stability and difficulties in large-scale growth limit its integration into devices. Conversely, p-type tellurium (Te) with excellent hole mobility upto 980 cm$^2$V$^{-1}$S$^{-1}$ \cite{tong2020stable}, serves as a promising alternative \cite{qiu2022resurrection, wang2018field}. A straightforward hydrothermal growth \cite{mo2002controlled} or thermal evaporation technique \cite{chen2007fabrication} can synthesize large-scale, high-quality tellurium with good air stability. Additionally, Te exhibits both 2D and 1D structures, controllable via simple temperature variation in vapor deposition processes \cite{zhao2022controllable}. Recent studies have reported the development of various 2D/2D p-n junctions, which hold significant applications in next-generation functional devices \cite{bernardi2013extraordinary, lee2014atomically, xia2022pristine}, including junctions such as WSe$_2$/MoS$_2$ \cite{cheng2014electroluminescence, li2015epitaxial} and MoTe$_2$/MoS$_2$ \cite{duong2019modulating, paul2017photo}, among others. In contrast, 1D materials, which offer nearly defect-free channels, are highly suitable for nano-electronics due to their well-defined linear geometry and directional conductivity. Additionally, the antenna effect of nanowires enhances their optoelectronic properties, making them excellent candidates for 1D/2D heterostructures \cite{meng2024anti,jadwiszczak2022mixed, lu2021reconfigurable,wang2022mixed, wang2022highly}.

  We report a vapour deposition based sequential growth of MoS$_2$ flakes followed by Te nanowire (NW), provides a pathway to create mixed-dimensional intigrated circuits. Most vdWs heterostructures reported to date have been created through a process involving micromechanical exfoliation and manual restacking. By utilizing this growth technique, it is possible to realise the device performances of both individual 1D and 2D materials, as well as their heterojunctions.Micro-Raman spectroscopy, scanning electron microscopy (SEM), atomic force microscopy (AFM), and Kelvin probe force microscopy (KPFM) were employed for the preliminary characterization of the intersecting junction and its individual sample counterparts. We find type-I like band alignment which exhibits conventional diode-like transport characteristics at the junction.
  To validate our transport data, we conducted current-voltage measurements using conductive AFM (CAFM). 
  Density functional theory (DFT) combined with non equilibrium Green’s function calculations were employed to calculate local device density of states (LDDOS) and the corresponding band bending, which support our experimental data. Using the non heterostructure 1D and 2D sample regions, we performed individual FET operations on Si/SiO$_2$, whose performances were significantly enhanced by using ionic liquid (DEME TFSI) gating.
  Finally, we implemented low power-based complementary metal oxide semiconductor (CMOS) and basic logic gate applications using the Te NW/MoS$_2$ pass transistor logic.

\section{RESULTS AND DISCUSSIONS}
\begin{figure*}[h!]
    \centerline{\includegraphics[scale=0.8, clip]{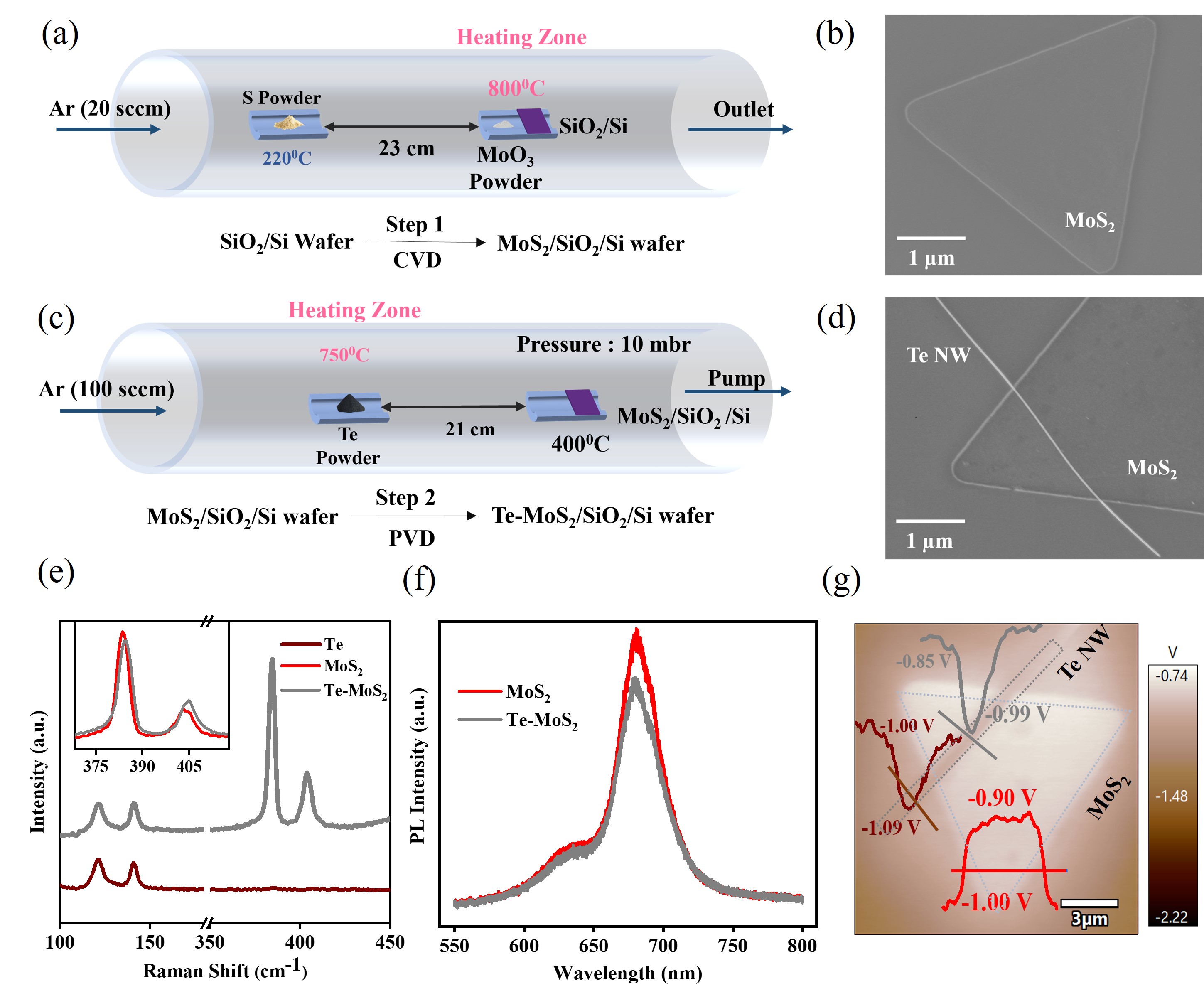}}
    \caption{Schematic of growth setup and characterisation of as-grown Te-MoS$_2$ heterostructures. (a) Schematic of CVD setup for the growth of MoS$_2$. (b) SEM image of CVD grown MoS$_2$. (c) Schematic of PVD setup for the growth of Te NW on MoS$_2$ grown substrate. (d) SEM image of Te-MOs$_2$ heterostructure. (e) Raman spectra of Te, MoS$_2$ and overlapping areas in the heterostructure. Brown line shows the Raman spectra of Te NW and that of the operlapping region is shown by the gray line and that of pristine MoS$_2$ is shown by red line. Inset shows the up shift of both Raman charecteristics peaks in overlapped MoS$_2$ from pristine MoS$_2$. (f) Photoluminescence spectra of monolayer MoS$_2$ from the pristine flake region (red) and the heterojunction region (gray). (g) Mapping of the surface potential of the heterostructure using KPFM to obtain the work function of Te NW (brown line-cut), MoS$_2$ (red line-cut) and overlapping region (gray line-cut).} 
    \label{fig1}
\end{figure*}

 \textbf{Growth of MoS$_2$, Te NW and their heterojunctions.} A two step vapor deposition process was used for the large scale  growth of MoS$_2$ and Te NW. Initially, an atmospheric pressure CVD technique was adopted to grow MoS$_2$ flakes on SiO$_2$ (285 nm)/Si substrate \cite{zhou2018unveiling}. A schematic of the single-furnace CVD setup is shown in Figure \ref{fig1} (a). Blank SiO$_2$/Si substrates (1 cm × 1 cm) were placed on an alumina boat containing molybdenum oxide (MoO$_3$) powder (2 mg) at the center of the furnace. Sulfur (S) powder (200 mg) was placed in a different boat at upstream side, 23 cm away from the center of the MoO$_3$ boat. Subsequently, the quartz tube was flushed using 100 sccm (standard cubic centimeters per minute) flow of Ar gas for 15 minutes. The furnace temperature was raised to 800$^{\circ}$C at a rate of 12$^{\circ}$C/min and maintained for 15 min with a flow of 20 sccm of Ar as carrier gas. Finally, the furnace was cooled down to room temperature naturally, and the substrates were unloaded. After the successful growth of MoS$_2$, the substrates were dipped in isopropyl alcohol (IPA) for 3 min. Physical vapor deposition (PVD) technique was used to grow Te NWs on the same MoS$_2$ grown substrates \cite{wang2014van}. For this, an alumina boat containing Te (5 mg) powder was placed in the center of the furnace, while the MoS$_2$ grown SiO$_2$/Si were placed on a different boat, 21 cm downstream from the center (Figure \ref{fig1} (c)). The temperature of the center zone was raised to 750$^{\circ}$C at a rate of 16$^{\circ}$C/min and the temperature was maintained for 5 minutes, with a flow of 100 sccm of Ar. The pressure inside the tube was maintained at 10 mbar throughout the growth time using a pump. Finally, the system was cooled down to ambient temperature. After the complete two step fabrication process, the final substrate consists of a dense assembly of 1D and 2D structures along with their intersecting heterojunctions, which are taken for further experiment. Typically, in a $130 \mu m \times 130 \mu m$ area, 20-25 Te NWs, 50-60 MoS$_2$ flakes and 15-20 Te NW/MoS$_2$ heterojunctions can be counted (see Figure S1). The heterostructures are selected for the fabrication of the vdW diodes. On the other hand, the individual NW and MoS$_2$ flakes spread throughout the wafer can be considered as the building blocks for the individual FETs. Thus, we can fabricate an in-situ CVD/PVD template of discrete Te NW/MoS$_2$ (p-n) on a large scale of several centimeters (see Figure S1), enabling the operation of p-n heterostructures and various types of FET applications. The morphology of the 1D Te/2D MoS$_2$ heterosrtucture with good uniformity was confirmed by SEM image as shown in Figure $\ref{fig1}$ (d). The approximate height of the MoS$_2$ flake and diameter of the Te NW are 2 nm and 8 nm, respectively as measured by tapping mode AFM (see the corresponding line cut of the height profile in Figure S2).

\textbf{Characterisation of MoS$_2$, Te NW and their heterojunctions.}
Figure \ref{fig1} (e) shows the Raman spectra of the as-grown MoS$_2$ (red), Te NW/MoS$_2$ overlapping region (gray), and Te NW (brown). Two distinct phonon peaks of Te NW are observed in the Raman spectra. The peak situated at $\sim$ 121.4 cm$^{-1}$ is related to the A$_1$ mode, due to the chain-expansion vibration of Te atoms along the basal plane \cite{pine1971raman}. Another peak, at $\sim$ 140.9 cm$^{-1}$, corresponds to the E$_2$ mode, consistent with previous reports \cite{qin2020raman,du2017one}. In case of MoS$_2$, the positions of both characteristic modes show a finite red shift, as shown in the inset of the Figure \ref{fig1} (e). The red shift for both modes suggests electron transfer from MoS$_2$ to the NW \cite{maguire2019defect}. To investigate the interfacial coupling between Te NW and MoS$_2$, we performed photoluminescence (PL) measurements at the heterostructure and pristine MoS$_2$ as shown in Figure \ref{fig1} (f). The PL peak for MoS$_2$ appears at $\sim$ 680 nm (1.82 eV), and its intensity slightly diminishes at the heterojunction due to nonradiative recombination caused by charge transfer \cite{jadwiszczak2022mixed}.

Local surface potential mapping along the Te NW/MoS$_2$ junction is carried out using KPFM, as described in Figure \ref{fig1} (g). The contrast differences of the KPFM-micrograph imply the change in surface potential on different regions on the substrate: (i) SiO$_2$/Te, (ii) Te/MoS$_2$, and (iii) SiO$_2$/MoS$_2$. Three line cuts and their associated surface potential profiles in region (i), (ii) and (iii) are indicated by the brown, grey and red lines, respectively in the Figure \ref{fig1} (g). The results reveal that MoS$_2$ has higher surface potential than both Te ($\Delta V_{CPD}$ = 0.14) and SiO$_2$ ($\Delta V_{CPD}$ = -0.10), while Te has lower ($\Delta V_{CPD}$ = 0.09) work function than SiO$_2$. Hence, the diffusion of the electrons takes place across the interface to form the depletion region with a built-in potential (V$_{bi}$). The work function of individual Te and MoS$_2$ can be obtained from the measured surface potential difference ($\Delta$V$_{CPD}$) using the following equations, $ \Delta V_{CPD} = \frac{1}{e}(W_{Substrate} - W_{Sample}) $, where e is the electronic charge, W$_{Substrate}$ and W$_{Sample}$ are the work function of the SiO$_2$ substrate and the sample, respectively. Considering $W_{substrate}=5.05$ eV \cite{lee2009interlayer}, the work function of the Te NW and MoS$_2$ are calculated to be 4.96 eV and 5.15 eV, respectively. The observed work function values of MoS$_2$ and Te are consistent with previous reports \cite{choi2014layer,michaelson1977work}. Moreover, the difference between the Fermi levels ($\Delta$F$_F$) of Te and MoS$_2$ is $ 0.19$ eV. Due to this, electrons transfer from MoS$_2$ to Te until thermal equilibrium is established, forming a type-I band alignment with conduction band offset, $\Delta$E$_C$ $= $ 0.36 eV and valance band offset, $\Delta$E$_V$ $= $ 1.09 eV (see Figure S3).

\begin{figure*}[h!]
    \centerline{\includegraphics[scale=0.8, clip]{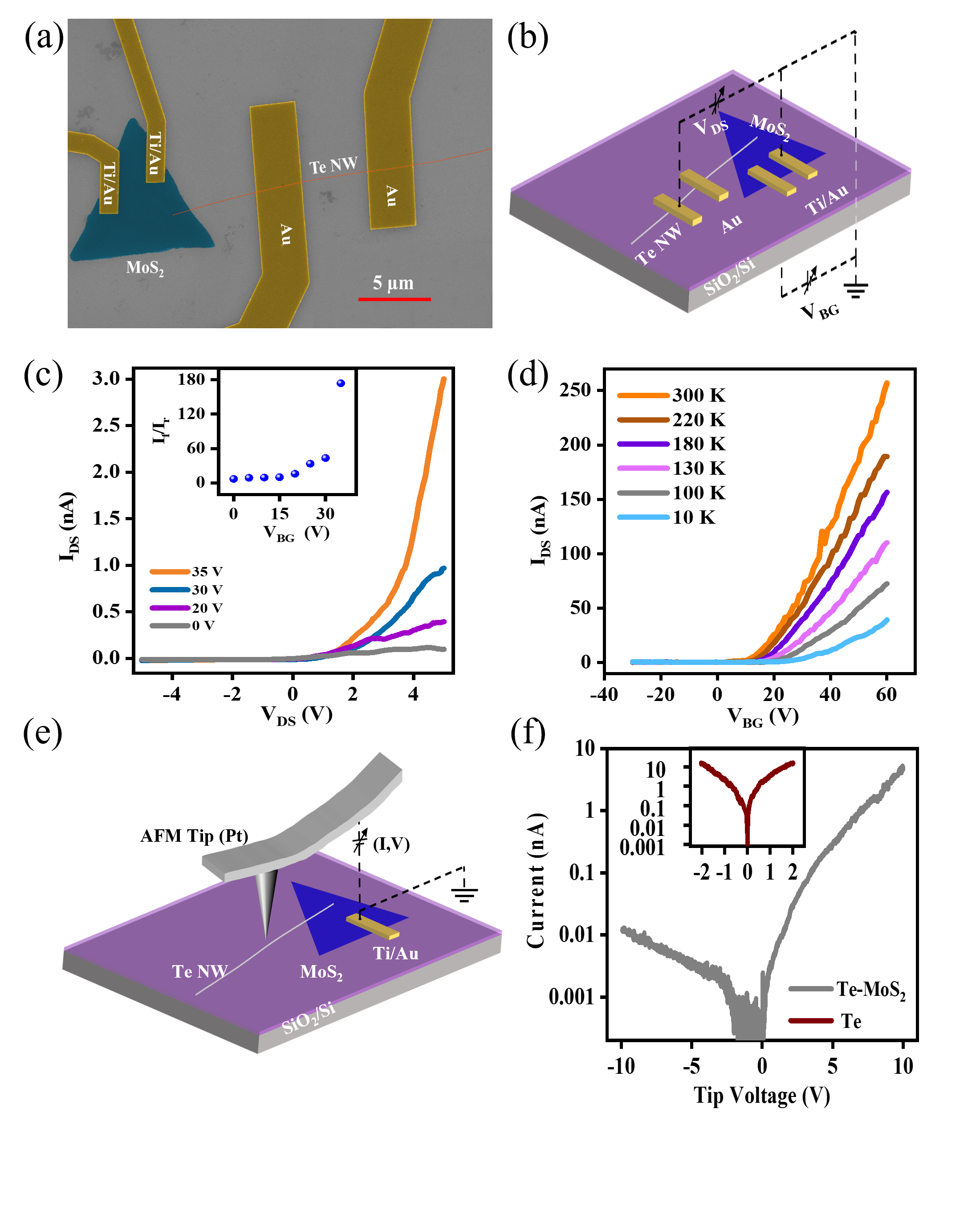}}
    \caption{Electrical transport measurement on SiO$_2$. (a) False-colored SEM image of a tasted device fabricated by electron beam lithography (EBL).  
    (b) Schematic of mixed-channel diode device. (c) I-V curves at different gate voltages measured across the p-n heterojunction device as shown in the schematic (b) (drain at Te NW and source at MoS$_2$). Inset shows the rectification ratio at different gate voltages. (d) Transfer characteristics of p-n heterojunction FET at V$_{DS}$ = 3V at different temperatures. (e) Schematic representation of the CAFM for lateral transport measurement across Te/MoS$_2$ heterostructure. Voltage sweeps from 0 V up to $\pm$ 10 V were applied to the tip (MoS$_2$ is grounded), and the current through the sample was simultaneously measured. The I–V measurements were carried out with 20 nA current compliance. (f) The semi-log scale plot of current−voltage characteristic measured by CAFM across Te (brown) and the heterojunction (gray).} 
    \label{fig2}
\end{figure*}

\textbf{Electrical transport of Te NW/MoS$_2$ heterojunction diode.}
A false color SEM image of a representative heterostructure device is provided in Figure $\ref{fig2}$ (a). 
The device structure consists of a Te NW FET, a Te NW/MoS$_2$ mixed dimensional dual-channel FET, and a MoS$_2$ FET for the independent electrical characterization (representative images are shown in Figure S4 (a) and (b)). During the FET measurements, a back-gate voltage (V$_{BG}$) was swept from -60 to +60 V. Figure S5(a) represents the transfer characteristics curves for both the devices. Expected p-type (black) and n-type (blue) transfer curves are obtained for Te NW and MoS$_2$ FETs, respectively.

The mobility of the Te NW FET is calculated using the following relation from ref \cite{ford2009diameter} applicable for a 1D channel, $\mu_p = g_m\frac{L^2}{C_{ox}}\frac{1}{V_{DS}}$, where L = 5 $\mu$m is the channel length, V$_{DS}$ = 0.1 V is the applied drain-source bias, $g_m (= \frac{dI_{DS}}{dV_{BG}}$|$_{V_{DS}}$) is the transconductance and C$_{ox}$ is the gate capacitance of a cylindrical nanowire on a planar substrate. The value of C$_{ox}$ is calculated using the following expression, $C_{ox} = \frac{2\pi\epsilon\epsilon_0L}{cosh^{-1}[(r+t_{ox})/r]}$, where dielectric constant of the gate insulator ($\epsilon$ for SiO$_2$) is 3.9, r = 20 nm is the radius of the Te NW, $t_{ox}$ = 285 nm is the thickness of the dielectric medium and $\epsilon_0$ the permittivity of free space. The low field field-effect mobility of the Te NW is found to be $\sim$ 734 cm$^2$V$^{-1}$s$^{-1}$, calculated from the above mentioned mobility equation. The device has an on-off ratio $\sim$ 6.4 (see Figure S5(a)). Similarly, the mobility of MoS$_2$ FET is found to be $\sim$ 8.9 cm$^2$V$^{-1}$s$^{-1}$ using a slightly different formula applicable for 2D channels, $\mu_n = \frac{dI_{DS}}{dV_{BG}} \frac{L}{WC_iV_{DS}}$, from the reference \cite{radisavljevic2011single}, Where L = 2 µm is the channel length, W = 5 $\mu$m is the channel width, C$_i$ = 1.2$\times$10 $^{-8}$ Fcm$^{-2}$ is the capacitance between the channel and the gate per unit area ( $C_i= \frac{\epsilon \epsilon_0}{t_{ox}}$ ), V$_{DS}=0.5$ V. The device has an on-off ratio $\sim$ $1.8 \times 10^6$. 

The donor and acceptor concentration (N$_D$ and N$_A$) for the n and p- channels are calculated by employing the following expressions, $N_D = \frac{I_{DS}L}{ e \mu_n V_{DS} t W}$ and $N_A = \frac{I_{DS}L}{ \pi e \mu_p V_{DS} r^2}$, from the reference \cite{wang2022mixed}, where t is the thickness of MoS$_2$. The channel current I$_{DS}$ is extracted from the corresponding transfer curve at V$_{BG}=0$ V. The obtained values of the doping concentrations are $N_A=1.05 \times 10^{19}$ $cm^{-3}$ and $N_D=2.36 \times 10^{18}$ $cm^{-3}$. We estimate the depletion width in the p-n junction using the following expressions, $ W_{Te}= \sqrt{\frac{2N_D\epsilon_0\epsilon_a\epsilon_dV_{bi}}{eN_A(\epsilon_aN_A+\epsilon_dN_D)}}$ and $W_{MoS_2}=\sqrt{\frac{2N_A\epsilon_0\epsilon_a\epsilon_dV_{bi}}{eN_D(\epsilon_aN_A+\epsilon_dN_D)}}$, where $\epsilon_0$ is the permittivity of free space. $\epsilon_a$ and $\epsilon_d$ are the dielectric constants of Te and MoS$_2$ respectively ($\epsilon_a$ $\sim$ 4.12 \cite{sharma2022structural}, $\epsilon_d \sim$ 4 \cite{santos2013electrically}). The calculated values of W$_{Te}$ and W$_{MoS_2}$ are 5.59 nm and 1.26 nm, respectively. Hence, the total depletion width is $W_{total}=W_{Te}+W_{MoS_2}=$ 6.85 nm. A narrow depletion region indicates a large internal electric field ($2.92 \times 10^7 Vm^{-1}$) develops across the interface, which is consistent with the previous reports on few atomic layers p-n junctions \cite{chu2018atomic,wang2022mixed}.

The output curves of the individual channel materials show linear behaviour, which can be seen in Figure S5 (b). Interestingly, the output curves  of the heterojunction display conventional p-n junction diode behavior when Te is kept at positive bias and MoS$_2$ is connected with the ground (see Figure $\ref{fig2}$ (c)). The forward bias current increases with rising V$_{BG}$, while the reverse saturation current remains nearly constant. The similar transport behaviour has been confirmed for other five to six heterostructure devices. We observed the knee voltage ($V_{knee}$) is $\sim$ 0.14 V at zero back-gate bias, which is almost consistent with our KPFM data. The gate tunable current rectification ratio (I$_f$/I$_r$) (due to change in band alignment) is shown in the inset of Figure \ref{fig2} (c) which is explained in later section.  

The ideality factor (n) of the diode is estimated to be $\sim$ 2.60 (at zero applied back-gate voltage) at room temperature using the following expression, $n = \frac{e}{k_BT}\frac{dV}{dlnI}$, where I is the forward current, V is the applied voltage, T is the temperature in Kelvin, e is electronic charge and k$_B$ is the Boltzmann’s constant.

Next, we investigated the temperature dependent (10-300 K) three-terminal measurement on the mixed-channel p-n FET, keeping the source terminal at MoS$_2$ and the drain terminal at Te NW, as shown in Figure $\ref{fig2}$ (b). During the measurement, the back-gate voltage was swept from -30 V to +60V, and the drain-source volatge was fixed at 3 V. The heterojunction exhibited n-type transport behaviour throughout the entire temperature range (see Figure $\ref{fig2}$ (d)) due to electron dominated transport in the MoS$_2$ channel \cite{yao2021high}. At room temperature, the on-off ratio of this device is $\sim$ 250, which decreases with temperature. For better realization of the transport data, electronic structure calculations of Te/MoS$_2$ heterostructure are performed. Figure S6 (a) shows the representative heterostructure. The heterostructure exhibits a semiconducting behavior with band gap $\sim$ 0.5 eV and type-I band alignment, which has potential applications in device electronics. The calculated orbital-projected density of states of the heterostructure are depicted in Figure S6 (b). Orbital analysis illustrates that the conduction band minimum (CBM) and valence band maximum (VBM) are predominantly contributed by Mo-4d orbital and Te-5p orbital, respectively. As the conduction bands near the Fermi level are mostly contributed by MoS$_2$, it evidently manifests the n-type transport in Te/MoS$_2$ heterojunction. The redistribution of charge between Te and MoS$_2$ layers were also investigated. Bader charge analysis resembles that electrons are mainly transferred from Te to MoS$_2$ layer. MoS$_2$ gains 0.0028 e- per Te atom from Te. The interlayer electron redistribution enhances the performance of Te by reducing the band gap of heterojunction and increases the carrier mobility as well.

 Room temperature current-voltage (I-V) measurements were also performed using CAFM, with Ti/Au and Au serving as the contacts to MoS$_2$ and Te NW, respectively \cite{huang2018van}, as indicated in the Figure $\ref{fig2}$ (e). A conducting tip was used to measure the local current on different positions of the individual semiconductors and heterostructures. The Te NW/MoS$_2$ heterostructure exhibits diode-like feature (tip voltage $\pm$ 10 V), yielding a current rectification ratio of around 400, as shown in Figure $\ref{fig2}$ (f). On Te NW alone, (see Figure S7) we observed a symmetric I-V characteristic (depicted in the inset of Figure $\ref{fig2}$ (f)). Furthermore, this p-n junction diode possesses an ideality factor of n = 3.15, calculated using the above equation of ideality factor. We were unable to measure the current on the individual MoS$_2$ flakes by CAFM because the measured current was near the experimental resolution limit of 0.1 pA.

 The current rectification ratio (I$_f$/I$_r$), which compares forward current (I$_f$) and reverse currents (I$_f$), is plotted against V$_{BG}$ for the p-n heterojunction diode, as given in the inset of Figure \ref{fig2} (c). A noticeable change in the rectification ratio occurs as V$_{BG}$ swept from -20 to 30 V. The variation in the I$_f$/I$_r$ ratio arises from changes in the band structures of each semiconductor and alterations in the band alignment at the hetero-interface. Specifically, the n-type MoS$_2$ undergoes a transition from a nearly insulating intrinsic (\textit{i}) state at V$_{BG}$ = -20 V to a heavily n-doped (n$^+$) state at V$_{BG}$ = 30 V. Te NW is a low band gap p-type semiconductor, the Fermi level shifts slightly away from the valence band as the gate voltage varies from negative to positive. Consequently, the p-type Te NW undergoes an opposite transition from a heavily p-doped (p$^+$) state at V$_{BG}$ = -20 V to a lightly p-doped state (p$^-$) at V$_{BG}$ = 30 V. At V$_{BG}$ = 0 V, a p-n junction forms between Te NW/MoS$_2$, which transforms into a p$^-$/n$^+$ junction (Figure S8 (b)) as V$_{BG}$ increases further (Figure S8 (c)). In this p$^-$/n$^+$ state, the shift of the Fermi level towards the conduction band is greater in MoS$_2$ than Te. Consequently, both channels of the heterojunction are in the "on" state under positive gate bias. The forward bias current increases with rising gate voltage as the carrier concentration in MoS$_2$ increases. When forward bias is applied to the heterojunction, it disrupts the equilibrium state of the junction, leading to a decrease in the built-in potential (with the external electric field opposing the built-in electrical field). Conversely, reverse bias enhances the built-in potential, hindering the movement of majority carriers. This increased built-in field generates a small drift current of minority carriers. However, minority carriers of Te (electrons) cannot overcome the conduction band offset, while minority carriers (holes) of MoS$_2$ must pass through $\Delta$E$_V$, resulting in a constant reverse saturation current. Moreover, the MoS$_2$ channel is in the "off" state when a negative V$_{BG}$ is applied, forming a p$^+$/\textit{i} junction (Figure S8 (a)). Consequently, the movement of charge carriers through the MoS$_2$ channel is restricted due to its high resistance, leading to a gradual decrease in forward bias current as the gate voltage becomes negative. This behavior explains the observed low current rectification ratio at negative V$_{BG}$. The diode's highest rectification ratio is 180 at V$_{DS}$ =4, which is comparable to other similar studies \cite{wang2022highly}. 

\begin{figure*}[h!]
    \centerline{\includegraphics[scale=1, clip]{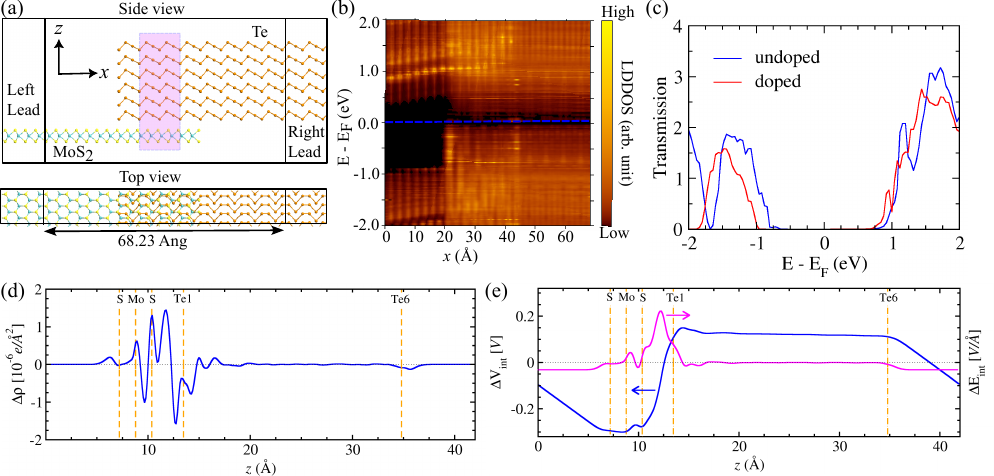}}
    \caption{(a) Top and side views of a two-terminal device consisting of a monolayer MoS$_2$ and a six-layer Te vdWs heterojunction, with the monolayer MoS$_2$ and six layers of Te wire as the left lead (LL) and right lead RL), respectively, used for quantum transport simulations. Green, yellow, and orange solid spheres denote the atomic positions of Mo, S, and Te atoms, respectively. (b) The position- and energy-dependent local device density of states (LDDOS) across the transport channel as show in panel(a). (c) Transmission spectrum for charged neutral (black line) and doped (blue line) device set-up. For the charge-doped case, MoS$_2$ and Te are doped with n-type and p-type carriers, respectively, with a carrier density of 10$^{-14}$ cm$^{-3}$. Panels (d) and (e) show the differential xy-averaged charge density, $\Delta\rho$, and electrostatic potential, $\Delta$V, respectively, along the z-axis in the overlapped region as indicated by the pink shaded box in panel (a) for a pristine set-up. Orange dashed lines indicate the positions of the lower S, Mo, upper S, interfacial Te(Te1), and outermost Te(Te6) layers from the interface. Magenta line in panel indicate the electric field $\Delta$E$_{int}$ = $\delta$(V$_{int}$)/$\delta$z:}
    \label{Theory}
\end{figure*}

\textbf{Theoretical results.}
In order to get insights on the charge transport through Te/MoS$_2$ vdW heterojunction, we have carried out first-principle based quantum transport calculations (see experimental method for computational details) on the two terminal device structures as illustrated in Figure \ref{Theory} (a). An effective n-type and p-type doping can be simulated by incorporating appropriate compensation charges \cite{stradi2016general,soler2002siesta} into the MoS$_2$ and Te systems, respectively, mimicking the experimentally observed intrinsic doping in the individual layers. This effective doping scheme, utilizing atomic compensation charge, offers significant advantages as it remains unaffected by the system’s exact dimensions or the specific characteristics of the dopant atoms \cite{stradi2016general}. Here, we consider experimentally relevant doping densities of 10$^{14}$ cm$^{-3}$ for both n-type and p-type.

The band alignment of the doped heterostructure is depicted in the energy-position resolved local device density of states (LDDOS), as shown in Figure \ref{Theory} (b). Interestingly, band bending occurs at the conduction band of the overlapped junctions of the heterostructure, whereas the valence band remains almost unaffected. Additionally, significant band bending does not occur in the Te layer because its DFT bandgap ( 0.3 eV) is much smaller than that of the monolayer MoS$_2$  (1.8 eV).

Figure \ref{Theory} (c) illustrates the zero bias transmission spectra of the Te/MoS$_2$  vdW heterojunction both for the charge doped and undoped device set-up. In the doped case, the excess negative and positive charge carriers in the MoS$_2$ and Te layers, respectively, result in strong interlayer coupling and cause an almost rigid shift of the transmission spectrum towards the positive energy. Notably, this polarity of the excess charges due to intrinsic doping on the individual layers follows the same polarity as the charge transfer at the pristine heterojunction, i.e., electrons and holes accumulated in the MoS$_2$ and Te layers, respectively as shown in Figure \ref{Theory} (d). The corresponding differential built-in electrostatic potential, $\Delta$V$_{int}$ and eletric field, E$_{int}$ across the heteojunction is dipected in Figure \ref{Theory} (e). DFT calculated V$_{int}$ is about 0.4 V at the vdW gap of the heterojunction, which is close to the experimentally estimated built-in potential of approximately 0.19 V.

\textbf{FET characteristics of Te NW and MoS$_2$ with IL gate.}
Next, the performances of the individual Te NW and MoS$_2$ FETs were examined using IL (DEME-TFSI) as a top gate. The measurements were carried out in a vacuum chamber to eliminate any electrochemical degradation and air exposure. The IL was dropped on the channel materials, covering the source, drain and gate electrodes. The IL gate voltage is swept at a rate of 10 mV/sec to avoid hysteresis caused by the slow-moving ions in the IL. The maximum gate voltage is limited to ±2V to prevent any potential reactions between the ions and the channel materials. The superiority of IL over other solid-state gate dielectric is the formation of an electric double layer (EDL) between the channel and IL which acts as a narrow-gap capacitor with a significantly higher capacitance. Hence, the injected charge density is an order of magnitude higher than that of the conventional SiO$_{2}$ back-gate \cite{wang2015ionic,lieb2019ionic}. This surface charge density can tune structural transitions and induce superconductivity, ferromagnetism, and other correlated phenomena in various classes of materials \cite{jo2015electrostatically,chien2016thermoelectric,yamada2011electrically,nakano2012collective,zhang2019electric}. When a positive voltage is applied to the IL gate electrode, negative ions in the IL accumulate near the gate electrode, while positive ions gather near the MoS$_2$ channel. When a negative voltage is applied to the gate, the movement of ions reverses. In both scenarios, electric double layers form at the interfaces between the IL and the solid surfaces.

\begin{figure*}[h!]
    \centerline{\includegraphics[scale=0.6, clip]{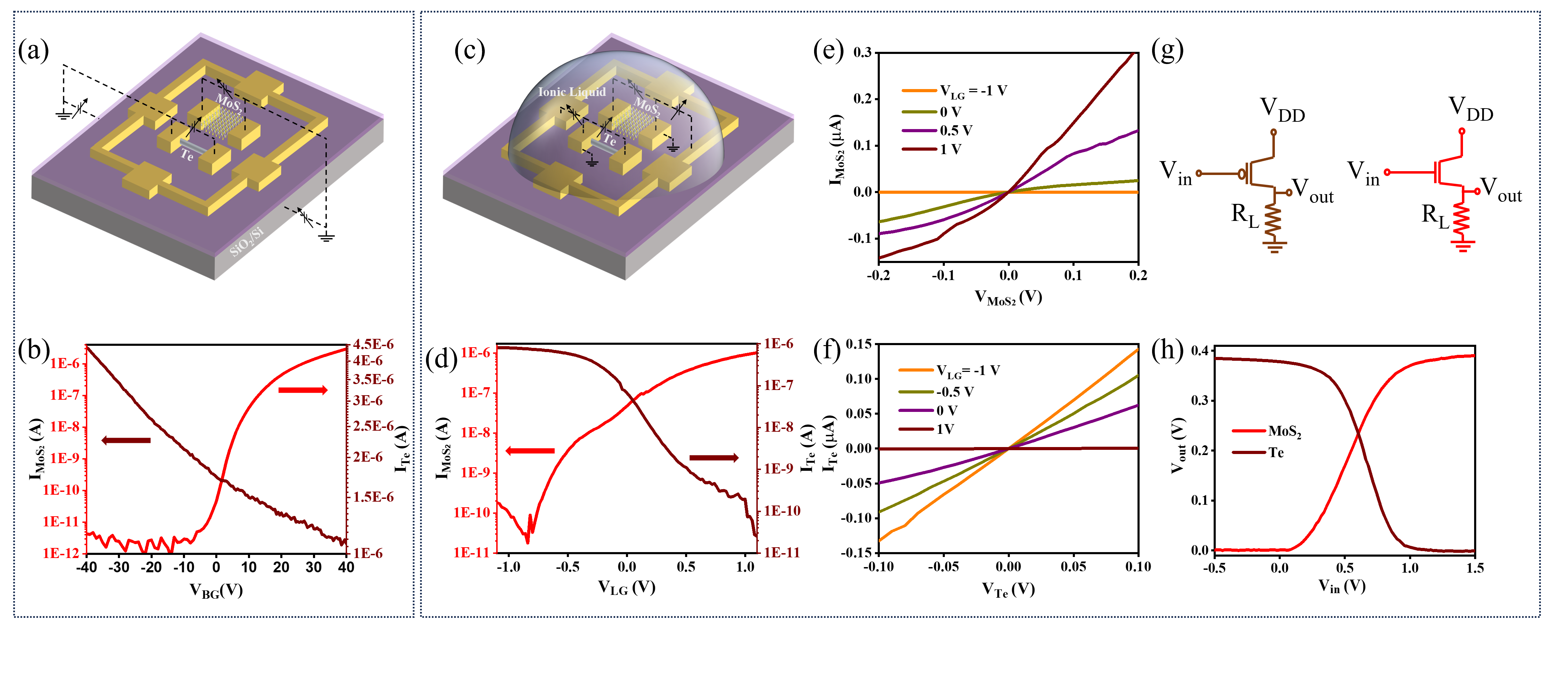}}
    \caption{
    (a) Schematic of an SiO$_2$ gated Te NW and MoS$_2$ FET. (b) Transconductance curve of Te NW (brown) and MoS$_2$ (red) with SiO$_2$ back-gate. (c) Schematic of an IL-gated Te NW and MoS$_2$ FET. (d) Transconductance curve of Te NW (brown) and MoS$_2$ (red). (e) and (f) I-V characteristics of MoS$_2$ and Te devices measured at different IL gate voltages, respectively. (f) Circuit diagrams of of unipolar logic gate (Schmitt trigger) of Te NW (brown) and MoS$_2$ (red) with series load resistances. (g) Output curves of Schmitt trigger devices (Te: brown, MoS$_2$: red).} 
    \label{fig3}
\end{figure*}

Figure \ref{fig3} (a) and (c) illustrates the schematic of the FET measurements for the same Te nanowire and MoS$_2$ devices with SiO$_2$ and IL gating, respectively. The devices are measured using both the gates to compare their performances. The surrounding Au pads serve as the gate electrodes for the IL. Figure \ref{fig3} (b) shows the transfer characteristics curve of Te and MoS$_2$ FETs with SiO$_2$ as back-gate. The transfer characteristics display the same behavior as previously observed for the individual channel of the heterostructure, as explained earlier (Figure \ref{fig2} (a) and Figure S5 (a)). The mobility and on-off ratio of this particular NW FET are determined to be $\sim$ 468 cm$^2V^{-1}s^{-1}$ and $\sim$ 4.1, respectively (r = 12 nm, L = 4 $\mu$m). Similarly for MoS$_2$ the parameters are $\sim$ 8 cm$^2V^{-1}s^{-1}$ and $\sim$ $2.4 \times 10^6$, respectively (L = 4 $\mu$m, W = 5 $\mu$m).

After applying the IL, at $V_{LG}=0$, the drain-source current for both devices is modified (see Figure \ref{fig3} (d)) compared to the current observed with SiO$_2$ back-gate. At $V_{LG}=0$, the drain-source current for Te became one order of magnitude lower compared to the current with $SiO_{2}$ as back-gate. Te is a heavily doped p-type material and does not possess any proper on-off state during $SiO_{2}$ gate. Additionally, Te provides nearly contact-barrier free FET as reported by Dasika et al. \cite{dasika2021contact} Therefore, adding the IL to Te NW may increase the Schottky barrier height at the junctions, leading to a lower drain-source current compared to the bare Te NW on $SiO_{2}$. Contrarily, for MoS$_2$, the drain-source current at $V_{LG}=0$ is 2 orders of magnitude higher, which can be attributed to the initial charge transfer from ionic liquid, resulting in surface doping \cite{lieb2019ionic}. The EDL capacitance can be determined using the relation: $\frac{\Delta V_{BG}}{\Delta V_{LG}}=\frac{C_{BG}}{C_{LG}}$, where $\Delta V_{BG}$ and $\Delta V_{LG}$ represent the gate voltage differences required to achieve the same change in conductivity in the linear region of the transfer curves for back-gate and ionic liquid gate applications, respectively \cite{chu2014charge}. For MoS$_2$, the EDL capacitnace ($C_{LG}$) was calculated to be $\sim$ $9.6 \times 10^{-6}$ $F$cm$^{-2}$. Consequently, the mobility was determined to be $\sim$ 42 cm$^2V^{-1}s^{-1}$. Due to the lower saturation current value observed with IL gating compared to the current across the entire back-gate window, we could not use this comparison relation to calculate the mobility of the IL-gated Te NW. The on-off ratios for both Te NW and MoS$_2$ FET are observed to be $\sim$ $5.1 \times 10^{4}$ and $\sim$ $5.4 \times 10^{4}$, respectively, with IL gating. The higher off current (20 pA) compared to SiO$_2$ back-gate (Figure \ref{fig3} (d)) is attributed to the high leakage current of the ionic liquid itself (0.1 nA at 1 V) for both channels. Nevertheless, IL gate is significantly more effective in enhancing the overall performance (such as mobility improvement and low-voltage operation) of both devices compared to the SiO$_2$ back-gate. Moreover the I-V characteristics for both MoS$_2$ and Te NW are measured at different IL gate as shown in Figure \ref{fig3} (e) and (f) respectively. The non-linear I-V curves for MoS$_2$ indicates the presence of a Schottky barrier at the metal-semiconductor junction, while for Te, the behaviour is Ohmic. Unipolar logic with a series connection of load resistance (Schmitt trigger) are also constructed using individual Te and MoS$_2$ FET. Figure \ref{fig3} (f) describes the circuit diagram and voltage transfer characteristics of unipolar logic using MOS$_2$ (red) and Te FET (brown), respectively. The applied V$_{DD}$ = 0.4 volt. The output voltages are measured across a 10 M$\Omega$ load resistance. Figure \ref{fig3} (f) shows the voltage transfer curve of unipolar logic which toggles between `0' and `1' depending on the input gate voltage for both the cases.

\begin{figure*}[h!]
    \centerline{\includegraphics[scale=0.8, clip]{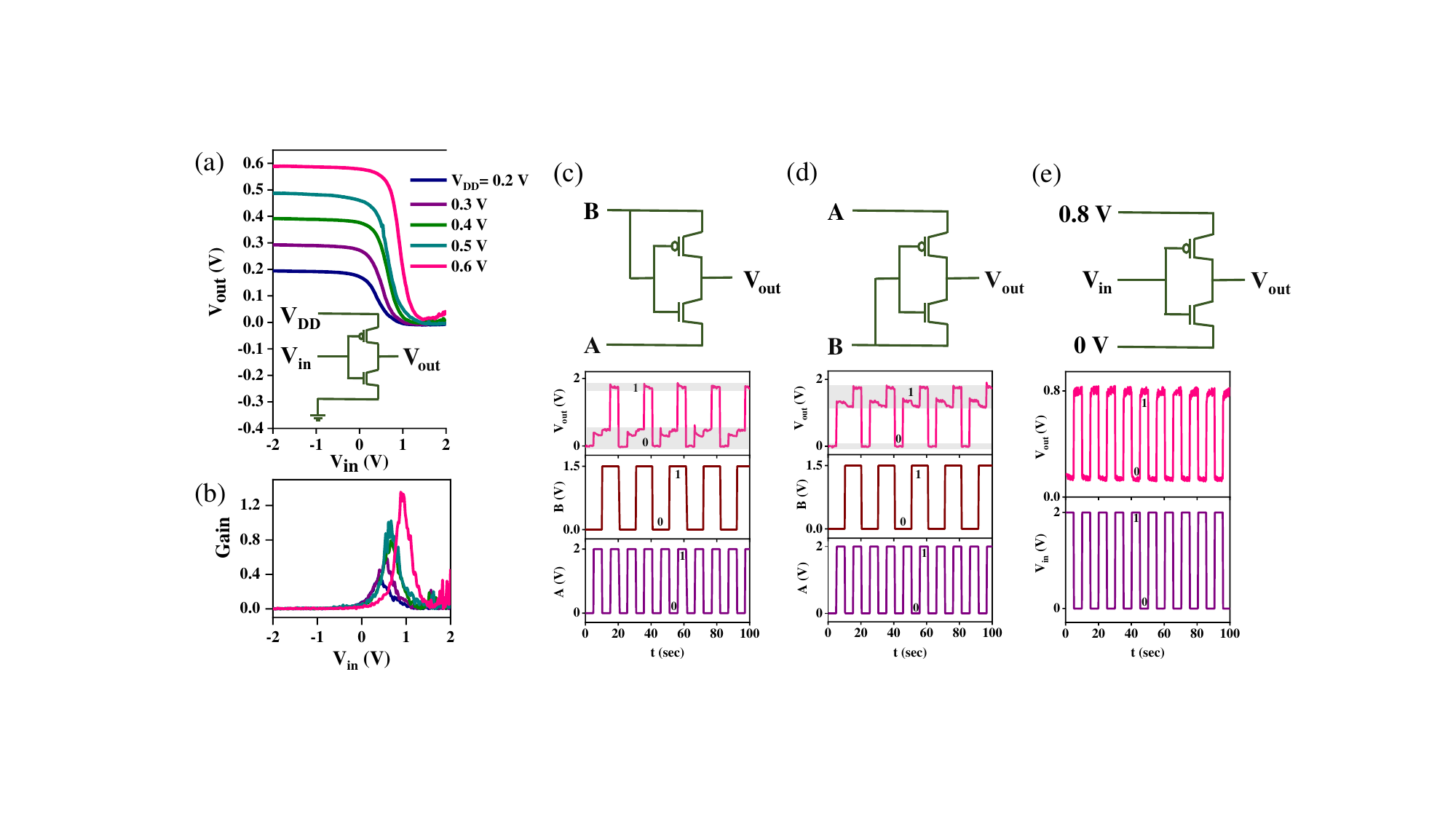}}
    \caption{CMOS and logic-gates using ionic-liquid gating. (a)  Voltage transfer (V$_{in}$-V$_{out}$) CMOS inverter at different V$_{DD}$. Inset shows the schematic of electrical configuration for the CMOS inverter. (b) Gain (-dV$_{out}$/dV$_{in}$) characteristics of the CMOS inverter at different V$_{DD}$. (c), (d) and (e) provides the schematic circuit diagram and dynamic output voltage response of AND, OR and NOT gate, respectively. } 
    \label{fig4}
\end{figure*}

\textbf{CMOS and basic logic operations with IL gating.}
An IL gate-based CMOS inverter is designed by connecting n-type MoS$_2$ and p-type Te FETs, both of which exhibit good transport characteristics. The inset of  Figure \ref{fig4} (a) shows the circuit diagram of the CMOS, where Te and MoS$_2$ FETs are connected in series. The output voltage is measured across the MoS$_2$ FET, whereas, the input voltage is applied to the common gate terminals.  Figure \ref{fig4} (a) illustrates the voltage transfer characteristics of the 1D/2D CMOS, where a steep transition between the two logic states (0 and 1) is observed at different applied $V_{DD}$ values. The gains of the CMOS at different $V_{DD}$ are plotted in Figure \ref{fig4} (b) The maximum gain ($-\frac{dV_{out}}{dV_{in}}$) is calculated to be 1.2 with a maximum static DC power of 10 nW, makes it a low-power CMOS suitable for practical applications. After successfully realizing the heterostructure CMOS, we implemented the basic logic gate operations using the same CMOS-based device components. AND, OR, and NOT gates were constructed using the pass transistor circuit model \cite{ding2012cmos}. The circuit diagrams for AND, OR and NOT gates along with their corresponding input vs. output signals, are shown in \ref{fig4} (c), (d) and (e) respectively. The two inputs, A and B, were toggled between two voltage states of 0 V and 1.5 V (2 V for NOT gate), and the observed output states perfectly followed the truth tables of the corresponding basic gates.

\section{CONCLUSIONS} 
 In summary, we have described the synthesis of a mixed-dimensional (1D/2D) vdWs assembly of p-type Te NWs and n-type MoS$_2$ and their heterojunctions using a two-step vapor deposition technique. This method for growing in-situ vdW heterostructures have rarely been applied in diode applications so far. The grown samples have dual applications: (a) using the 1D/2D heteointerface, one can study a $<10$ nm thin depletion region, i.e., an atomically thin diode, and (b) realizing individual FET and digital logic operations through their combinations.  
 The morphology of the individual sample is primarily characterized by optical microscopy, SEM, and AFM. 
 The charge transfer across the interface has been confirmed by micro-Raman spectroscopy. KPFM studies confirm a type-I band alignment with a built-in potential of 0.19 V along the heterostructure. Electrical transport studies show that the device functions as a gate-tunable p-n junction diode with a maximum rectification ratio $\sim$180 due to a very high electric field generated across the interface. Similarly, CAFM studies confirm the diode-like behavior. The ideality factor, depletion width, and gate tunable retification ratio have been calculated from the mixed-channel FET characterization. First-principle based quantum transport calculations on the two terminal device confirm the possible band bending. Moreover, the ionic liquid gating enhances the performance of Te FET, while the performance of MoS$_2$ FET is not affected much. The contrasting IL gate-tunable performance of these two FET helps us to obtain similar on-off ratio, which is a key to design CMOS.  Consequently, the low-power CMOS inverter and basic logic gate operations have been realized using the IL gated p- and n-FET combinations on a single substrate.
 
 Our growth technique offers a potential pathway to create high density, high-quality, defect-free interfaces, which enables the design and fabrication of next-generation functional devices. However, the properties of junction capacitance and the potential use of these atomically thin vdWs diodes as rectifiers, along with the suitability of thier FET counterpart for advanced digital electronic applications, still need to be explored. Additionally, the optoelectronic properties combined with field tunability may offer opportunities in broader applications.

 \section{METHODS}

\textbf{Characterization.} The individual Te NWs and MoS$_2$ flakes were primarily examined under an optical microscope (OLYMPUS BX53M) and SEM (JEOL JSM-6010LA). The heights of the nanostructures were confirmed by tapping mode AFM (Asylum Research MFP-3D). Surface potential mapping was performed using kelvin probe force microscopy (KPFM) equipped with the same AFM setup.Micro  Raman spectroscopy and micro photoluminescence (PL) measurement was conducted using the Horiba T64000 raman spectrometer (excitation wavelength of 532 nm with a spot size of $\sim$ 1$\mu$m) at different positions of a heterostructure.

\textbf{Device fabrication and electrical measurements.} Electrical contacts on the heterostructures and individual p and n channels were made using optical lithography/electron beam lithography (EBL) (Zeiss Sigma 300 with Raith Elphy Quantum), followed by metallization inside a thermal evaporator. Au (80 nm) and Ti/Au (8 nm/80 nm) were deposited on Te NWs and MoS$_2$ flakes, respectively, to ensure Ohmic contact at the metal-semiconductor junctions. Local gate electrodes were also fabricated during the lithography process. Electrical transport  measurements (two and three terminal) were carried out using Keithley 2450 source measure units (SMU), Keithley 2601B SMU and Keithley 2182A Nanovoltmeter. Low temperature transport measurements were performed inside a closed cycle cryostat (ARS-4HW). For the lateral transport measurements of the devices, CAFM technique was employed, using a Ti/Ir-coated (5 nm/20 nm) silicon tip from Asylum Research. A schematic of the CAFM experiment is shown in Figure \ref{fig2} (g). 

\textbf{Ionic liquid gate.} The performances of the individual FETs, heterostructure CMOS, and logic gates were examined using IL (DEME-TFSI, sigma aldrich) gating as an better alternative to SiO$_2$. A small drop of ionic liquid (IL) was applied using a micropipette onto the MoS$_2$ and Te channels, covering the source, drain, and gate electrodes, inside an argon filled glove box. After drop-casting the IL, the devices were instantly transferred to the measurement chamber to ensure the minimum air-exposure. To avoid any type of electrochemical reactions, all transport measurements involving the IL were conducted at a vacuum of $10^{-6}$ mbar and at room temperature.

\textbf{Computational Methodology.} Electronic structure calculations of Te/MoS$_2$ heterostructure were performed in density functional theory framework using Vienna ab initio simulation package (VASP) \cite{kresse1996efficient,kresse1996efficiency}. Generalized gradient approximation (GGA) with the Perdew-Burke-Ernzerhof (PBE) \cite{perdew1996generalized} realization for the exchange-correlation potential was employed and the electron-ion interaction was described through the projector augmented wave (PAW) method. Interlayer vdW interaction was implemented by DFT-D2 method developed by Grimme. The plane-wave cutoff energy was set to be 380 eV. Ionic relaxation of the heterostructure was carried out by the conjugate-gradient scheme until the Hellmann-Feynman forces and energy differences were achieved up to $10^{-3}$ eV/Å and $10^{-5}$ eV, respectively. Gamma centered k-mesh of 8×8×6 and 12×12×8 for lattice relaxation and density of states calculations, respectively were used to integrate the Brillouin zone.

We employ the interface builder \cite{QuantumATK2023} in the QuantumATK package \cite{smidstrup2019quantumatk} to construct an Te/monolayer-MoS$_2$ heterostructure. Quantum transport calculations for the heterostructure are carried out within the framework of density functional theory (DFT) combined with nonequllibirum Green’s function (NEGF) formalism using the Perdew-Burke-Ernzerhof (PBE) parametrization \cite{perdew1996generalized} of the generalized gradient approximation (GGA) to the exchange-correlation functional, as implemented in QuantumATK package \cite{smidstrup2019quantumatk}. Norm-conserving fully relativistic pseudopotentials of the type PseudoDojo-SO \cite{smidstrup2019quantumatk, van2018pseudodojo} for describing electron-core interactions; and the Pseudojojo (medium) numerical linear combination of atomic orbitals (LCAO) basis set \cite{van2018pseudodojo}. The energy mesh cutoff for the real-space grid is chosen as 101 Hartree, and the k-point grid $161$ is used for the self-consistent calculations. Local density of states and transmission are obtained by integrating over $151$ k-point mesh. Periodic boundary condition is used for the self-consistent calculations, and a 15 $\AA$ vacuum is placed on the top of heterostructure in order to remove interaction between the consecutive periodic image.
\section{ACKNOWLEDGMENTS}
  BK and BD would like to thank Mr. Sanjib Naskar, Mr. Soham Ash for experimental helps and analysis. BK, BD and MP are grateful to IACS for fellowships. SMaity and TK are grateful to DST-INSPIRE for their fellowships. RP is grateful to CSIR for his fellowship. SM and SDas are greatful to UGC for fellowship. KD is greatful to TRC for his fellowship. SD acknowledges the financial support from DST-SERB grant No. SCP/2022/000411 and CRG/2021/004334. SD acknowledges the CSS facilities and support from the Technical Research Centre (TRC), IACS.

\section{Author Contributions}
BK and BD contributed equally to the project. SD supervised the project. SD and BD conceived the project. BK and BD synthesized the samples taking the help from RP. BK performed the heterojunction diode based measurements and analysis. BD did the IL based CMOS and logic gate operations. BK did the basic characterisations with BD, SM, RP, SDas, TK and KD. SMaity and MP fabricated the device structures. Theoretical calculations were performed by TK and KDolui. Manuscript is written by BK and BD taking the inputs from SM, TK, KDolui and KD. The final draft is corrected and modified by SD.

\section{Conflict of Interest}
The authors declare no competing interests.

\section{Keywords}
 Te NW, MoS$_2$, vapor deposition, vdW p-n junction, digital electronics.

\bibliography{Ref}

\begin{thebibliography}{70}%
\makeatletter
\providecommand \@ifxundefined [1]{%
 \@ifx{#1\undefined}
}%
\providecommand \@ifnum [1]{%
 \ifnum #1\expandafter \@firstoftwo
 \else \expandafter \@secondoftwo
 \fi
}%
\providecommand \@ifx [1]{%
 \ifx #1\expandafter \@firstoftwo
 \else \expandafter \@secondoftwo
 \fi
}%
\providecommand \natexlab [1]{#1}%
\providecommand \enquote  [1]{``#1''}%
\providecommand \bibnamefont  [1]{#1}%
\providecommand \bibfnamefont [1]{#1}%
\providecommand \citenamefont [1]{#1}%
\providecommand \href@noop [0]{\@secondoftwo}%
\providecommand \href [0]{\begingroup \@sanitize@url \@href}%
\providecommand \@href[1]{\@@startlink{#1}\@@href}%
\providecommand \@@href[1]{\endgroup#1\@@endlink}%
\providecommand \@sanitize@url [0]{\catcode `\\12\catcode `\$12\catcode
  `\&12\catcode `\#12\catcode `\^12\catcode `\_12\catcode `\%12\relax}%
\providecommand \@@startlink[1]{}%
\providecommand \@@endlink[0]{}%
\providecommand \url  [0]{\begingroup\@sanitize@url \@url }%
\providecommand \@url [1]{\endgroup\@href {#1}{\urlprefix }}%
\providecommand \urlprefix  [0]{URL }%
\providecommand \Eprint [0]{\href }%
\providecommand \doibase [0]{https://doi.org/}%
\providecommand \selectlanguage [0]{\@gobble}%
\providecommand \bibinfo  [0]{\@secondoftwo}%
\providecommand \bibfield  [0]{\@secondoftwo}%
\providecommand \translation [1]{[#1]}%
\providecommand \BibitemOpen [0]{}%
\providecommand \bibitemStop [0]{}%
\providecommand \bibitemNoStop [0]{.\EOS\space}%
\providecommand \EOS [0]{\spacefactor3000\relax}%
\providecommand \BibitemShut  [1]{\csname bibitem#1\endcsname}%
\let\auto@bib@innerbib\@empty
\bibitem [{\citenamefont {Reis}\ \emph {et~al.}(2006)\citenamefont {Reis},
  \citenamefont {Lubaszewski},\ and\ \citenamefont {Jess}}]{reis2006design}%
  \BibitemOpen
  \bibfield  {author} {\bibinfo {author} {\bibfnamefont {R.}~\bibnamefont
  {Reis}}, \bibinfo {author} {\bibfnamefont {M.}~\bibnamefont {Lubaszewski}},\
  and\ \bibinfo {author} {\bibfnamefont {J.}~\bibnamefont {Jess}},\ }\href@noop
  {} {\emph {\bibinfo {title} {Design of systems on a chip: design and test}}}\
  (\bibinfo  {publisher} {Springer},\ \bibinfo {year} {2006})\BibitemShut
  {NoStop}%
\bibitem [{\citenamefont {Cao}\ \emph {et~al.}(2023)\citenamefont {Cao},
  \citenamefont {Bu}, \citenamefont {Vinet}, \citenamefont {Cao}, \citenamefont
  {Takagi}, \citenamefont {Hwang}, \citenamefont {Ghani},\ and\ \citenamefont
  {Banerjee}}]{Cao2023}%
  \BibitemOpen
  \bibfield  {author} {\bibinfo {author} {\bibfnamefont {W.}~\bibnamefont
  {Cao}}, \bibinfo {author} {\bibfnamefont {H.}~\bibnamefont {Bu}}, \bibinfo
  {author} {\bibfnamefont {M.}~\bibnamefont {Vinet}}, \bibinfo {author}
  {\bibfnamefont {M.}~\bibnamefont {Cao}}, \bibinfo {author} {\bibfnamefont
  {S.}~\bibnamefont {Takagi}}, \bibinfo {author} {\bibfnamefont
  {S.}~\bibnamefont {Hwang}}, \bibinfo {author} {\bibfnamefont
  {T.}~\bibnamefont {Ghani}},\ and\ \bibinfo {author} {\bibfnamefont
  {K.}~\bibnamefont {Banerjee}},\ }\href@noop {} {\bibfield  {journal}
  {\bibinfo  {journal} {Nature}\ }\textbf {\bibinfo {volume} {621}},\ \bibinfo
  {pages} {1476} (\bibinfo {year} {2023})}\BibitemShut {NoStop}%
\bibitem [{\citenamefont {Akinwande}\ \emph {et~al.}(2019)\citenamefont
  {Akinwande}, \citenamefont {Huyghebaert}, \citenamefont {Wang}, \citenamefont
  {Serna}, \citenamefont {Goossens}, \citenamefont {Li}, \citenamefont {Wong},\
  and\ \citenamefont {Koppens}}]{akinwande2019graphene}%
  \BibitemOpen
  \bibfield  {author} {\bibinfo {author} {\bibfnamefont {D.}~\bibnamefont
  {Akinwande}}, \bibinfo {author} {\bibfnamefont {C.}~\bibnamefont
  {Huyghebaert}}, \bibinfo {author} {\bibfnamefont {C.-H.}\ \bibnamefont
  {Wang}}, \bibinfo {author} {\bibfnamefont {M.~I.}\ \bibnamefont {Serna}},
  \bibinfo {author} {\bibfnamefont {S.}~\bibnamefont {Goossens}}, \bibinfo
  {author} {\bibfnamefont {L.-J.}\ \bibnamefont {Li}}, \bibinfo {author}
  {\bibfnamefont {H.-S.~P.}\ \bibnamefont {Wong}},\ and\ \bibinfo {author}
  {\bibfnamefont {F.~H.}\ \bibnamefont {Koppens}},\ }\href@noop {} {\bibfield
  {journal} {\bibinfo  {journal} {Nature}\ }\textbf {\bibinfo {volume} {573}},\
  \bibinfo {pages} {507} (\bibinfo {year} {2019})}\BibitemShut {NoStop}%
\bibitem [{\citenamefont {Pendurthi}\ \emph {et~al.}(2024)\citenamefont
  {Pendurthi}, \citenamefont {Sakib}, \citenamefont {Sadaf}, \citenamefont
  {Zhang}, \citenamefont {Sun}, \citenamefont {Chen}, \citenamefont
  {Jayachandran}, \citenamefont {Oberoi}, \citenamefont {Ghosh}, \citenamefont
  {Kumari} \emph {et~al.}}]{pendurthi2024monolithic}%
  \BibitemOpen
  \bibfield  {author} {\bibinfo {author} {\bibfnamefont {R.}~\bibnamefont
  {Pendurthi}}, \bibinfo {author} {\bibfnamefont {N.~U.}\ \bibnamefont
  {Sakib}}, \bibinfo {author} {\bibfnamefont {M.~U.~K.}\ \bibnamefont {Sadaf}},
  \bibinfo {author} {\bibfnamefont {Z.}~\bibnamefont {Zhang}}, \bibinfo
  {author} {\bibfnamefont {Y.}~\bibnamefont {Sun}}, \bibinfo {author}
  {\bibfnamefont {C.}~\bibnamefont {Chen}}, \bibinfo {author} {\bibfnamefont
  {D.}~\bibnamefont {Jayachandran}}, \bibinfo {author} {\bibfnamefont
  {A.}~\bibnamefont {Oberoi}}, \bibinfo {author} {\bibfnamefont
  {S.}~\bibnamefont {Ghosh}}, \bibinfo {author} {\bibfnamefont
  {S.}~\bibnamefont {Kumari}}, \emph {et~al.},\ }\href@noop {} {\bibfield
  {journal} {\bibinfo  {journal} {Nature nanotechnology}\ ,\ \bibinfo {pages}
  {1}} (\bibinfo {year} {2024})}\BibitemShut {NoStop}%
\bibitem [{\citenamefont {Jayachandran}\ \emph {et~al.}(2024)\citenamefont
  {Jayachandran}, \citenamefont {Pendurthi}, \citenamefont {Sadaf},
  \citenamefont {Sakib}, \citenamefont {Pannone}, \citenamefont {Chen},
  \citenamefont {Han}, \citenamefont {Trainor}, \citenamefont {Kumari},
  \citenamefont {Mc~Knight} \emph {et~al.}}]{jayachandran2024three}%
  \BibitemOpen
  \bibfield  {author} {\bibinfo {author} {\bibfnamefont {D.}~\bibnamefont
  {Jayachandran}}, \bibinfo {author} {\bibfnamefont {R.}~\bibnamefont
  {Pendurthi}}, \bibinfo {author} {\bibfnamefont {M.~U.~K.}\ \bibnamefont
  {Sadaf}}, \bibinfo {author} {\bibfnamefont {N.~U.}\ \bibnamefont {Sakib}},
  \bibinfo {author} {\bibfnamefont {A.}~\bibnamefont {Pannone}}, \bibinfo
  {author} {\bibfnamefont {C.}~\bibnamefont {Chen}}, \bibinfo {author}
  {\bibfnamefont {Y.}~\bibnamefont {Han}}, \bibinfo {author} {\bibfnamefont
  {N.}~\bibnamefont {Trainor}}, \bibinfo {author} {\bibfnamefont
  {S.}~\bibnamefont {Kumari}}, \bibinfo {author} {\bibfnamefont {T.~V.}\
  \bibnamefont {Mc~Knight}}, \emph {et~al.},\ }\href@noop {} {\bibfield
  {journal} {\bibinfo  {journal} {Nature}\ }\textbf {\bibinfo {volume} {625}},\
  \bibinfo {pages} {276} (\bibinfo {year} {2024})}\BibitemShut {NoStop}%
\bibitem [{\citenamefont {Xia}\ \emph {et~al.}(2022{\natexlab{a}})\citenamefont
  {Xia}, \citenamefont {Zong}, \citenamefont {Pan}, \citenamefont {Chen},
  \citenamefont {Zhou}, \citenamefont {Song}, \citenamefont {Tong},
  \citenamefont {Guo}, \citenamefont {Ma}, \citenamefont {Gou} \emph
  {et~al.}}]{xia2022wafer}%
  \BibitemOpen
  \bibfield  {author} {\bibinfo {author} {\bibfnamefont {Y.}~\bibnamefont
  {Xia}}, \bibinfo {author} {\bibfnamefont {L.}~\bibnamefont {Zong}}, \bibinfo
  {author} {\bibfnamefont {Y.}~\bibnamefont {Pan}}, \bibinfo {author}
  {\bibfnamefont {X.}~\bibnamefont {Chen}}, \bibinfo {author} {\bibfnamefont
  {L.}~\bibnamefont {Zhou}}, \bibinfo {author} {\bibfnamefont {Y.}~\bibnamefont
  {Song}}, \bibinfo {author} {\bibfnamefont {L.}~\bibnamefont {Tong}}, \bibinfo
  {author} {\bibfnamefont {X.}~\bibnamefont {Guo}}, \bibinfo {author}
  {\bibfnamefont {J.}~\bibnamefont {Ma}}, \bibinfo {author} {\bibfnamefont
  {S.}~\bibnamefont {Gou}}, \emph {et~al.},\ }\href@noop {} {\bibfield
  {journal} {\bibinfo  {journal} {Small}\ }\textbf {\bibinfo {volume} {18}},\
  \bibinfo {pages} {2107650} (\bibinfo {year}
  {2022}{\natexlab{a}})}\BibitemShut {NoStop}%
\bibitem [{\citenamefont {Liu}\ \emph {et~al.}(2023)\citenamefont {Liu},
  \citenamefont {Niu}, \citenamefont {Yang}, \citenamefont {Chen},
  \citenamefont {Lu}, \citenamefont {Liao}, \citenamefont {Lu}, \citenamefont
  {Lu},\ and\ \citenamefont {Li}}]{liu2023large}%
  \BibitemOpen
  \bibfield  {author} {\bibinfo {author} {\bibfnamefont {M.}~\bibnamefont
  {Liu}}, \bibinfo {author} {\bibfnamefont {J.}~\bibnamefont {Niu}}, \bibinfo
  {author} {\bibfnamefont {G.}~\bibnamefont {Yang}}, \bibinfo {author}
  {\bibfnamefont {K.}~\bibnamefont {Chen}}, \bibinfo {author} {\bibfnamefont
  {W.}~\bibnamefont {Lu}}, \bibinfo {author} {\bibfnamefont {F.}~\bibnamefont
  {Liao}}, \bibinfo {author} {\bibfnamefont {C.}~\bibnamefont {Lu}}, \bibinfo
  {author} {\bibfnamefont {N.}~\bibnamefont {Lu}},\ and\ \bibinfo {author}
  {\bibfnamefont {L.}~\bibnamefont {Li}},\ }\href@noop {} {\bibfield  {journal}
  {\bibinfo  {journal} {Advanced Electronic Materials}\ }\textbf {\bibinfo
  {volume} {9}},\ \bibinfo {pages} {2200722} (\bibinfo {year}
  {2023})}\BibitemShut {NoStop}%
\bibitem [{\citenamefont {Navarro-Moratalla}\ \emph {et~al.}(2016)\citenamefont
  {Navarro-Moratalla}, \citenamefont {Island}, \citenamefont {Manas-Valero},
  \citenamefont {Pinilla-Cienfuegos}, \citenamefont {Castellanos-Gomez},
  \citenamefont {Quereda}, \citenamefont {Rubio-Bollinger}, \citenamefont
  {Chirolli}, \citenamefont {Silva-Guill{\'e}n}, \citenamefont {Agra{\"\i}t}
  \emph {et~al.}}]{navarro2016enhanced}%
  \BibitemOpen
  \bibfield  {author} {\bibinfo {author} {\bibfnamefont {E.}~\bibnamefont
  {Navarro-Moratalla}}, \bibinfo {author} {\bibfnamefont {J.~O.}\ \bibnamefont
  {Island}}, \bibinfo {author} {\bibfnamefont {S.}~\bibnamefont
  {Manas-Valero}}, \bibinfo {author} {\bibfnamefont {E.}~\bibnamefont
  {Pinilla-Cienfuegos}}, \bibinfo {author} {\bibfnamefont {A.}~\bibnamefont
  {Castellanos-Gomez}}, \bibinfo {author} {\bibfnamefont {J.}~\bibnamefont
  {Quereda}}, \bibinfo {author} {\bibfnamefont {G.}~\bibnamefont
  {Rubio-Bollinger}}, \bibinfo {author} {\bibfnamefont {L.}~\bibnamefont
  {Chirolli}}, \bibinfo {author} {\bibfnamefont {J.~A.}\ \bibnamefont
  {Silva-Guill{\'e}n}}, \bibinfo {author} {\bibfnamefont {N.}~\bibnamefont
  {Agra{\"\i}t}}, \emph {et~al.},\ }\href@noop {} {\bibfield  {journal}
  {\bibinfo  {journal} {Nature communications}\ }\textbf {\bibinfo {volume}
  {7}},\ \bibinfo {pages} {11043} (\bibinfo {year} {2016})}\BibitemShut
  {NoStop}%
\bibitem [{\citenamefont {Radisavljevic}\ \emph {et~al.}(2011)\citenamefont
  {Radisavljevic}, \citenamefont {Radenovic}, \citenamefont {Brivio},
  \citenamefont {Giacometti},\ and\ \citenamefont
  {Kis}}]{radisavljevic2011single}%
  \BibitemOpen
  \bibfield  {author} {\bibinfo {author} {\bibfnamefont {B.}~\bibnamefont
  {Radisavljevic}}, \bibinfo {author} {\bibfnamefont {A.}~\bibnamefont
  {Radenovic}}, \bibinfo {author} {\bibfnamefont {J.}~\bibnamefont {Brivio}},
  \bibinfo {author} {\bibfnamefont {V.}~\bibnamefont {Giacometti}},\ and\
  \bibinfo {author} {\bibfnamefont {A.}~\bibnamefont {Kis}},\ }\href@noop {}
  {\bibfield  {journal} {\bibinfo  {journal} {Nature nanotechnology}\ }\textbf
  {\bibinfo {volume} {6}},\ \bibinfo {pages} {147} (\bibinfo {year}
  {2011})}\BibitemShut {NoStop}%
\bibitem [{\citenamefont {Jariwala}\ \emph {et~al.}(2014)\citenamefont
  {Jariwala}, \citenamefont {Sangwan}, \citenamefont {Lauhon}, \citenamefont
  {Marks},\ and\ \citenamefont {Hersam}}]{jariwala2014emerging}%
  \BibitemOpen
  \bibfield  {author} {\bibinfo {author} {\bibfnamefont {D.}~\bibnamefont
  {Jariwala}}, \bibinfo {author} {\bibfnamefont {V.~K.}\ \bibnamefont
  {Sangwan}}, \bibinfo {author} {\bibfnamefont {L.~J.}\ \bibnamefont {Lauhon}},
  \bibinfo {author} {\bibfnamefont {T.~J.}\ \bibnamefont {Marks}},\ and\
  \bibinfo {author} {\bibfnamefont {M.~C.}\ \bibnamefont {Hersam}},\
  }\href@noop {} {\bibfield  {journal} {\bibinfo  {journal} {ACS nano}\
  }\textbf {\bibinfo {volume} {8}},\ \bibinfo {pages} {1102} (\bibinfo {year}
  {2014})}\BibitemShut {NoStop}%
\bibitem [{\citenamefont {Jariwala}\ \emph {et~al.}(2017)\citenamefont
  {Jariwala}, \citenamefont {Marks},\ and\ \citenamefont
  {Hersam}}]{jariwala2017mixed}%
  \BibitemOpen
  \bibfield  {author} {\bibinfo {author} {\bibfnamefont {D.}~\bibnamefont
  {Jariwala}}, \bibinfo {author} {\bibfnamefont {T.~J.}\ \bibnamefont
  {Marks}},\ and\ \bibinfo {author} {\bibfnamefont {M.~C.}\ \bibnamefont
  {Hersam}},\ }\href@noop {} {\bibfield  {journal} {\bibinfo  {journal} {Nature
  materials}\ }\textbf {\bibinfo {volume} {16}},\ \bibinfo {pages} {170}
  (\bibinfo {year} {2017})}\BibitemShut {NoStop}%
\bibitem [{\citenamefont {Geim}\ and\ \citenamefont
  {Grigorieva}(2013)}]{geim2013van}%
  \BibitemOpen
  \bibfield  {author} {\bibinfo {author} {\bibfnamefont {A.~K.}\ \bibnamefont
  {Geim}}\ and\ \bibinfo {author} {\bibfnamefont {I.~V.}\ \bibnamefont
  {Grigorieva}},\ }\href@noop {} {\bibfield  {journal} {\bibinfo  {journal}
  {Nature}\ }\textbf {\bibinfo {volume} {499}},\ \bibinfo {pages} {419}
  (\bibinfo {year} {2013})}\BibitemShut {NoStop}%
\bibitem [{\citenamefont {Haigh}\ \emph {et~al.}(2012)\citenamefont {Haigh},
  \citenamefont {Gholinia}, \citenamefont {Jalil}, \citenamefont {Romani},
  \citenamefont {Britnell}, \citenamefont {Elias}, \citenamefont {Novoselov},
  \citenamefont {Ponomarenko}, \citenamefont {Geim},\ and\ \citenamefont
  {Gorbachev}}]{haigh2012cross}%
  \BibitemOpen
  \bibfield  {author} {\bibinfo {author} {\bibfnamefont {S.~J.}\ \bibnamefont
  {Haigh}}, \bibinfo {author} {\bibfnamefont {A.}~\bibnamefont {Gholinia}},
  \bibinfo {author} {\bibfnamefont {R.}~\bibnamefont {Jalil}}, \bibinfo
  {author} {\bibfnamefont {S.}~\bibnamefont {Romani}}, \bibinfo {author}
  {\bibfnamefont {L.}~\bibnamefont {Britnell}}, \bibinfo {author}
  {\bibfnamefont {D.~C.}\ \bibnamefont {Elias}}, \bibinfo {author}
  {\bibfnamefont {K.~S.}\ \bibnamefont {Novoselov}}, \bibinfo {author}
  {\bibfnamefont {L.~A.}\ \bibnamefont {Ponomarenko}}, \bibinfo {author}
  {\bibfnamefont {A.~K.}\ \bibnamefont {Geim}},\ and\ \bibinfo {author}
  {\bibfnamefont {R.}~\bibnamefont {Gorbachev}},\ }\href@noop {} {\bibfield
  {journal} {\bibinfo  {journal} {Nature materials}\ }\textbf {\bibinfo
  {volume} {11}},\ \bibinfo {pages} {764} (\bibinfo {year} {2012})}\BibitemShut
  {NoStop}%
\bibitem [{\citenamefont {Kang}\ \emph {et~al.}(2013)\citenamefont {Kang},
  \citenamefont {Tongay}, \citenamefont {Zhou}, \citenamefont {Li},\ and\
  \citenamefont {Wu}}]{kang2013band}%
  \BibitemOpen
  \bibfield  {author} {\bibinfo {author} {\bibfnamefont {J.}~\bibnamefont
  {Kang}}, \bibinfo {author} {\bibfnamefont {S.}~\bibnamefont {Tongay}},
  \bibinfo {author} {\bibfnamefont {J.}~\bibnamefont {Zhou}}, \bibinfo {author}
  {\bibfnamefont {J.}~\bibnamefont {Li}},\ and\ \bibinfo {author}
  {\bibfnamefont {J.}~\bibnamefont {Wu}},\ }\href@noop {} {\bibfield  {journal}
  {\bibinfo  {journal} {Applied Physics Letters}\ }\textbf {\bibinfo {volume}
  {102}} (\bibinfo {year} {2013})}\BibitemShut {NoStop}%
\bibitem [{\citenamefont {Mak}\ \emph {et~al.}(2010)\citenamefont {Mak},
  \citenamefont {Lee}, \citenamefont {Hone}, \citenamefont {Shan},\ and\
  \citenamefont {Heinz}}]{mak2010atomically}%
  \BibitemOpen
  \bibfield  {author} {\bibinfo {author} {\bibfnamefont {K.~F.}\ \bibnamefont
  {Mak}}, \bibinfo {author} {\bibfnamefont {C.}~\bibnamefont {Lee}}, \bibinfo
  {author} {\bibfnamefont {J.}~\bibnamefont {Hone}}, \bibinfo {author}
  {\bibfnamefont {J.}~\bibnamefont {Shan}},\ and\ \bibinfo {author}
  {\bibfnamefont {T.~F.}\ \bibnamefont {Heinz}},\ }\href@noop {} {\bibfield
  {journal} {\bibinfo  {journal} {Physical review letters}\ }\textbf {\bibinfo
  {volume} {105}},\ \bibinfo {pages} {136805} (\bibinfo {year}
  {2010})}\BibitemShut {NoStop}%
\bibitem [{\citenamefont {Splendiani}\ \emph {et~al.}(2010)\citenamefont
  {Splendiani}, \citenamefont {Sun}, \citenamefont {Zhang}, \citenamefont {Li},
  \citenamefont {Kim}, \citenamefont {Chim}, \citenamefont {Galli},\ and\
  \citenamefont {Wang}}]{splendiani2010emerging}%
  \BibitemOpen
  \bibfield  {author} {\bibinfo {author} {\bibfnamefont {A.}~\bibnamefont
  {Splendiani}}, \bibinfo {author} {\bibfnamefont {L.}~\bibnamefont {Sun}},
  \bibinfo {author} {\bibfnamefont {Y.}~\bibnamefont {Zhang}}, \bibinfo
  {author} {\bibfnamefont {T.}~\bibnamefont {Li}}, \bibinfo {author}
  {\bibfnamefont {J.}~\bibnamefont {Kim}}, \bibinfo {author} {\bibfnamefont
  {C.-Y.}\ \bibnamefont {Chim}}, \bibinfo {author} {\bibfnamefont
  {G.}~\bibnamefont {Galli}},\ and\ \bibinfo {author} {\bibfnamefont
  {F.}~\bibnamefont {Wang}},\ }\href@noop {} {\bibfield  {journal} {\bibinfo
  {journal} {Nano letters}\ }\textbf {\bibinfo {volume} {10}},\ \bibinfo
  {pages} {1271} (\bibinfo {year} {2010})}\BibitemShut {NoStop}%
\bibitem [{\citenamefont {Miao}\ \emph {et~al.}(2017)\citenamefont {Miao},
  \citenamefont {Xu}, \citenamefont {Li}, \citenamefont {Bowman}, \citenamefont
  {Zhang}, \citenamefont {Hu}, \citenamefont {Zhou},\ and\ \citenamefont
  {Wang}}]{miao2017vertically}%
  \BibitemOpen
  \bibfield  {author} {\bibinfo {author} {\bibfnamefont {J.}~\bibnamefont
  {Miao}}, \bibinfo {author} {\bibfnamefont {Z.}~\bibnamefont {Xu}}, \bibinfo
  {author} {\bibfnamefont {Q.}~\bibnamefont {Li}}, \bibinfo {author}
  {\bibfnamefont {A.}~\bibnamefont {Bowman}}, \bibinfo {author} {\bibfnamefont
  {S.}~\bibnamefont {Zhang}}, \bibinfo {author} {\bibfnamefont
  {W.}~\bibnamefont {Hu}}, \bibinfo {author} {\bibfnamefont {Z.}~\bibnamefont
  {Zhou}},\ and\ \bibinfo {author} {\bibfnamefont {C.}~\bibnamefont {Wang}},\
  }\href@noop {} {\bibfield  {journal} {\bibinfo  {journal} {Acs Nano}\
  }\textbf {\bibinfo {volume} {11}},\ \bibinfo {pages} {10472} (\bibinfo {year}
  {2017})}\BibitemShut {NoStop}%
\bibitem [{\citenamefont {Liu}\ \emph {et~al.}(2017)\citenamefont {Liu},
  \citenamefont {Qu}, \citenamefont {Li}, \citenamefont {Moon}, \citenamefont
  {Ahmed}, \citenamefont {Kim}, \citenamefont {Lee}, \citenamefont {Choi},
  \citenamefont {Cho}, \citenamefont {Hone} \emph
  {et~al.}}]{liu2017modulation}%
  \BibitemOpen
  \bibfield  {author} {\bibinfo {author} {\bibfnamefont {X.}~\bibnamefont
  {Liu}}, \bibinfo {author} {\bibfnamefont {D.}~\bibnamefont {Qu}}, \bibinfo
  {author} {\bibfnamefont {H.-M.}\ \bibnamefont {Li}}, \bibinfo {author}
  {\bibfnamefont {I.}~\bibnamefont {Moon}}, \bibinfo {author} {\bibfnamefont
  {F.}~\bibnamefont {Ahmed}}, \bibinfo {author} {\bibfnamefont
  {C.}~\bibnamefont {Kim}}, \bibinfo {author} {\bibfnamefont {M.}~\bibnamefont
  {Lee}}, \bibinfo {author} {\bibfnamefont {Y.}~\bibnamefont {Choi}}, \bibinfo
  {author} {\bibfnamefont {J.~H.}\ \bibnamefont {Cho}}, \bibinfo {author}
  {\bibfnamefont {J.~C.}\ \bibnamefont {Hone}}, \emph {et~al.},\ }\href@noop {}
  {\bibfield  {journal} {\bibinfo  {journal} {ACS nano}\ }\textbf {\bibinfo
  {volume} {11}},\ \bibinfo {pages} {9143} (\bibinfo {year}
  {2017})}\BibitemShut {NoStop}%
\bibitem [{\citenamefont {Srivastava}\ \emph {et~al.}(2019)\citenamefont
  {Srivastava}, \citenamefont {Hassan}, \citenamefont {Gebredingle},
  \citenamefont {Jung}, \citenamefont {Kang}, \citenamefont {Yoo},
  \citenamefont {Singh},\ and\ \citenamefont {Lee}}]{srivastava2019van}%
  \BibitemOpen
  \bibfield  {author} {\bibinfo {author} {\bibfnamefont {P.~K.}\ \bibnamefont
  {Srivastava}}, \bibinfo {author} {\bibfnamefont {Y.}~\bibnamefont {Hassan}},
  \bibinfo {author} {\bibfnamefont {Y.}~\bibnamefont {Gebredingle}}, \bibinfo
  {author} {\bibfnamefont {J.}~\bibnamefont {Jung}}, \bibinfo {author}
  {\bibfnamefont {B.}~\bibnamefont {Kang}}, \bibinfo {author} {\bibfnamefont
  {W.~J.}\ \bibnamefont {Yoo}}, \bibinfo {author} {\bibfnamefont
  {B.}~\bibnamefont {Singh}},\ and\ \bibinfo {author} {\bibfnamefont
  {C.}~\bibnamefont {Lee}},\ }\href@noop {} {\bibfield  {journal} {\bibinfo
  {journal} {ACS applied materials \& interfaces}\ }\textbf {\bibinfo {volume}
  {11}},\ \bibinfo {pages} {8266} (\bibinfo {year} {2019})}\BibitemShut
  {NoStop}%
\bibitem [{\citenamefont {Tong}\ \emph {et~al.}(2020)\citenamefont {Tong},
  \citenamefont {Huang}, \citenamefont {Wang}, \citenamefont {Ye},
  \citenamefont {Peng}, \citenamefont {An}, \citenamefont {Sun}, \citenamefont
  {Zhang}, \citenamefont {Yang}, \citenamefont {Li} \emph
  {et~al.}}]{tong2020stable}%
  \BibitemOpen
  \bibfield  {author} {\bibinfo {author} {\bibfnamefont {L.}~\bibnamefont
  {Tong}}, \bibinfo {author} {\bibfnamefont {X.}~\bibnamefont {Huang}},
  \bibinfo {author} {\bibfnamefont {P.}~\bibnamefont {Wang}}, \bibinfo {author}
  {\bibfnamefont {L.}~\bibnamefont {Ye}}, \bibinfo {author} {\bibfnamefont
  {M.}~\bibnamefont {Peng}}, \bibinfo {author} {\bibfnamefont {L.}~\bibnamefont
  {An}}, \bibinfo {author} {\bibfnamefont {Q.}~\bibnamefont {Sun}}, \bibinfo
  {author} {\bibfnamefont {Y.}~\bibnamefont {Zhang}}, \bibinfo {author}
  {\bibfnamefont {G.}~\bibnamefont {Yang}}, \bibinfo {author} {\bibfnamefont
  {Z.}~\bibnamefont {Li}}, \emph {et~al.},\ }\href@noop {} {\bibfield
  {journal} {\bibinfo  {journal} {Nature communications}\ }\textbf {\bibinfo
  {volume} {11}},\ \bibinfo {pages} {2308} (\bibinfo {year}
  {2020})}\BibitemShut {NoStop}%
\bibitem [{\citenamefont {Qiu}\ \emph {et~al.}(2022)\citenamefont {Qiu},
  \citenamefont {Charnas}, \citenamefont {Niu}, \citenamefont {Wang},
  \citenamefont {Wu},\ and\ \citenamefont {Ye}}]{qiu2022resurrection}%
  \BibitemOpen
  \bibfield  {author} {\bibinfo {author} {\bibfnamefont {G.}~\bibnamefont
  {Qiu}}, \bibinfo {author} {\bibfnamefont {A.}~\bibnamefont {Charnas}},
  \bibinfo {author} {\bibfnamefont {C.}~\bibnamefont {Niu}}, \bibinfo {author}
  {\bibfnamefont {Y.}~\bibnamefont {Wang}}, \bibinfo {author} {\bibfnamefont
  {W.}~\bibnamefont {Wu}},\ and\ \bibinfo {author} {\bibfnamefont {P.~D.}\
  \bibnamefont {Ye}},\ }\href@noop {} {\bibfield  {journal} {\bibinfo
  {journal} {npj 2D Materials and Applications}\ }\textbf {\bibinfo {volume}
  {6}},\ \bibinfo {pages} {17} (\bibinfo {year} {2022})}\BibitemShut {NoStop}%
\bibitem [{\citenamefont {Wang}\ \emph {et~al.}(2018)\citenamefont {Wang},
  \citenamefont {Qiu}, \citenamefont {Wang}, \citenamefont {Huang},
  \citenamefont {Wang}, \citenamefont {Liu}, \citenamefont {Du}, \citenamefont
  {Goddard~III}, \citenamefont {Kim}, \citenamefont {Xu} \emph
  {et~al.}}]{wang2018field}%
  \BibitemOpen
  \bibfield  {author} {\bibinfo {author} {\bibfnamefont {Y.}~\bibnamefont
  {Wang}}, \bibinfo {author} {\bibfnamefont {G.}~\bibnamefont {Qiu}}, \bibinfo
  {author} {\bibfnamefont {R.}~\bibnamefont {Wang}}, \bibinfo {author}
  {\bibfnamefont {S.}~\bibnamefont {Huang}}, \bibinfo {author} {\bibfnamefont
  {Q.}~\bibnamefont {Wang}}, \bibinfo {author} {\bibfnamefont {Y.}~\bibnamefont
  {Liu}}, \bibinfo {author} {\bibfnamefont {Y.}~\bibnamefont {Du}}, \bibinfo
  {author} {\bibfnamefont {W.~A.}\ \bibnamefont {Goddard~III}}, \bibinfo
  {author} {\bibfnamefont {M.~J.}\ \bibnamefont {Kim}}, \bibinfo {author}
  {\bibfnamefont {X.}~\bibnamefont {Xu}}, \emph {et~al.},\ }\href@noop {}
  {\bibfield  {journal} {\bibinfo  {journal} {Nature Electronics}\ }\textbf
  {\bibinfo {volume} {1}},\ \bibinfo {pages} {228} (\bibinfo {year}
  {2018})}\BibitemShut {NoStop}%
\bibitem [{\citenamefont {Mo}\ \emph {et~al.}(2002)\citenamefont {Mo},
  \citenamefont {Zeng}, \citenamefont {Liu}, \citenamefont {Yu}, \citenamefont
  {Zhang},\ and\ \citenamefont {Qian}}]{mo2002controlled}%
  \BibitemOpen
  \bibfield  {author} {\bibinfo {author} {\bibfnamefont {M.}~\bibnamefont
  {Mo}}, \bibinfo {author} {\bibfnamefont {J.}~\bibnamefont {Zeng}}, \bibinfo
  {author} {\bibfnamefont {X.}~\bibnamefont {Liu}}, \bibinfo {author}
  {\bibfnamefont {W.}~\bibnamefont {Yu}}, \bibinfo {author} {\bibfnamefont
  {S.}~\bibnamefont {Zhang}},\ and\ \bibinfo {author} {\bibfnamefont
  {Y.}~\bibnamefont {Qian}},\ }\href@noop {} {\bibfield  {journal} {\bibinfo
  {journal} {Advanced materials}\ }\textbf {\bibinfo {volume} {14}},\ \bibinfo
  {pages} {1658} (\bibinfo {year} {2002})}\BibitemShut {NoStop}%
\bibitem [{\citenamefont {Chen}\ \emph {et~al.}(2007)\citenamefont {Chen},
  \citenamefont {Lu}, \citenamefont {Nie}, \citenamefont {Zhang}, \citenamefont
  {Zhang}, \citenamefont {Dai}, \citenamefont {Gao}, \citenamefont {Kan},
  \citenamefont {Li},\ and\ \citenamefont {Zou}}]{chen2007fabrication}%
  \BibitemOpen
  \bibfield  {author} {\bibinfo {author} {\bibfnamefont {H.}~\bibnamefont
  {Chen}}, \bibinfo {author} {\bibfnamefont {H.}~\bibnamefont {Lu}}, \bibinfo
  {author} {\bibfnamefont {Y.}~\bibnamefont {Nie}}, \bibinfo {author}
  {\bibfnamefont {J.}~\bibnamefont {Zhang}}, \bibinfo {author} {\bibfnamefont
  {M.}~\bibnamefont {Zhang}}, \bibinfo {author} {\bibfnamefont
  {Q.}~\bibnamefont {Dai}}, \bibinfo {author} {\bibfnamefont {S.}~\bibnamefont
  {Gao}}, \bibinfo {author} {\bibfnamefont {S.}~\bibnamefont {Kan}}, \bibinfo
  {author} {\bibfnamefont {D.}~\bibnamefont {Li}},\ and\ \bibinfo {author}
  {\bibfnamefont {G.}~\bibnamefont {Zou}},\ }\href@noop {} {\bibfield
  {journal} {\bibinfo  {journal} {Physics Letters A}\ }\textbf {\bibinfo
  {volume} {362}},\ \bibinfo {pages} {61} (\bibinfo {year} {2007})}\BibitemShut
  {NoStop}%
\bibitem [{\citenamefont {Zhao}\ \emph {et~al.}(2022)\citenamefont {Zhao},
  \citenamefont {Shi}, \citenamefont {Yin}, \citenamefont {Dong}, \citenamefont
  {Zhang}, \citenamefont {Kang}, \citenamefont {Yu}, \citenamefont {Chen},
  \citenamefont {Li}, \citenamefont {Liu} \emph
  {et~al.}}]{zhao2022controllable}%
  \BibitemOpen
  \bibfield  {author} {\bibinfo {author} {\bibfnamefont {X.}~\bibnamefont
  {Zhao}}, \bibinfo {author} {\bibfnamefont {J.}~\bibnamefont {Shi}}, \bibinfo
  {author} {\bibfnamefont {Q.}~\bibnamefont {Yin}}, \bibinfo {author}
  {\bibfnamefont {Z.}~\bibnamefont {Dong}}, \bibinfo {author} {\bibfnamefont
  {Y.}~\bibnamefont {Zhang}}, \bibinfo {author} {\bibfnamefont
  {L.}~\bibnamefont {Kang}}, \bibinfo {author} {\bibfnamefont {Q.}~\bibnamefont
  {Yu}}, \bibinfo {author} {\bibfnamefont {C.}~\bibnamefont {Chen}}, \bibinfo
  {author} {\bibfnamefont {J.}~\bibnamefont {Li}}, \bibinfo {author}
  {\bibfnamefont {X.}~\bibnamefont {Liu}}, \emph {et~al.},\ }\href@noop {}
  {\bibfield  {journal} {\bibinfo  {journal} {IScience}\ }\textbf {\bibinfo
  {volume} {25}} (\bibinfo {year} {2022})}\BibitemShut {NoStop}%
\bibitem [{\citenamefont {Bernardi}\ \emph {et~al.}(2013)\citenamefont
  {Bernardi}, \citenamefont {Palummo},\ and\ \citenamefont
  {Grossman}}]{bernardi2013extraordinary}%
  \BibitemOpen
  \bibfield  {author} {\bibinfo {author} {\bibfnamefont {M.}~\bibnamefont
  {Bernardi}}, \bibinfo {author} {\bibfnamefont {M.}~\bibnamefont {Palummo}},\
  and\ \bibinfo {author} {\bibfnamefont {J.~C.}\ \bibnamefont {Grossman}},\
  }\href@noop {} {\bibfield  {journal} {\bibinfo  {journal} {Nano letters}\
  }\textbf {\bibinfo {volume} {13}},\ \bibinfo {pages} {3664} (\bibinfo {year}
  {2013})}\BibitemShut {NoStop}%
\bibitem [{\citenamefont {Lee}\ \emph {et~al.}(2014)\citenamefont {Lee},
  \citenamefont {Lee}, \citenamefont {Van Der~Zande}, \citenamefont {Chen},
  \citenamefont {Li}, \citenamefont {Han}, \citenamefont {Cui}, \citenamefont
  {Arefe}, \citenamefont {Nuckolls}, \citenamefont {Heinz} \emph
  {et~al.}}]{lee2014atomically}%
  \BibitemOpen
  \bibfield  {author} {\bibinfo {author} {\bibfnamefont {C.-H.}\ \bibnamefont
  {Lee}}, \bibinfo {author} {\bibfnamefont {G.-H.}\ \bibnamefont {Lee}},
  \bibinfo {author} {\bibfnamefont {A.~M.}\ \bibnamefont {Van Der~Zande}},
  \bibinfo {author} {\bibfnamefont {W.}~\bibnamefont {Chen}}, \bibinfo {author}
  {\bibfnamefont {Y.}~\bibnamefont {Li}}, \bibinfo {author} {\bibfnamefont
  {M.}~\bibnamefont {Han}}, \bibinfo {author} {\bibfnamefont {X.}~\bibnamefont
  {Cui}}, \bibinfo {author} {\bibfnamefont {G.}~\bibnamefont {Arefe}}, \bibinfo
  {author} {\bibfnamefont {C.}~\bibnamefont {Nuckolls}}, \bibinfo {author}
  {\bibfnamefont {T.~F.}\ \bibnamefont {Heinz}}, \emph {et~al.},\ }\href@noop
  {} {\bibfield  {journal} {\bibinfo  {journal} {Nature nanotechnology}\
  }\textbf {\bibinfo {volume} {9}},\ \bibinfo {pages} {676} (\bibinfo {year}
  {2014})}\BibitemShut {NoStop}%
\bibitem [{\citenamefont {Xia}\ \emph {et~al.}(2022{\natexlab{b}})\citenamefont
  {Xia}, \citenamefont {Luo}, \citenamefont {Wang}, \citenamefont {Wang},
  \citenamefont {Li}, \citenamefont {Wang}, \citenamefont {Xu}, \citenamefont
  {Chen}, \citenamefont {Zhou}, \citenamefont {Wang} \emph
  {et~al.}}]{xia2022pristine}%
  \BibitemOpen
  \bibfield  {author} {\bibinfo {author} {\bibfnamefont {H.}~\bibnamefont
  {Xia}}, \bibinfo {author} {\bibfnamefont {M.}~\bibnamefont {Luo}}, \bibinfo
  {author} {\bibfnamefont {W.}~\bibnamefont {Wang}}, \bibinfo {author}
  {\bibfnamefont {H.}~\bibnamefont {Wang}}, \bibinfo {author} {\bibfnamefont
  {T.}~\bibnamefont {Li}}, \bibinfo {author} {\bibfnamefont {Z.}~\bibnamefont
  {Wang}}, \bibinfo {author} {\bibfnamefont {H.}~\bibnamefont {Xu}}, \bibinfo
  {author} {\bibfnamefont {Y.}~\bibnamefont {Chen}}, \bibinfo {author}
  {\bibfnamefont {Y.}~\bibnamefont {Zhou}}, \bibinfo {author} {\bibfnamefont
  {F.}~\bibnamefont {Wang}}, \emph {et~al.},\ }\href@noop {} {\bibfield
  {journal} {\bibinfo  {journal} {Light: Science \& Applications}\ }\textbf
  {\bibinfo {volume} {11}},\ \bibinfo {pages} {170} (\bibinfo {year}
  {2022}{\natexlab{b}})}\BibitemShut {NoStop}%
\bibitem [{\citenamefont {Cheng}\ \emph {et~al.}(2014)\citenamefont {Cheng},
  \citenamefont {Li}, \citenamefont {Zhou}, \citenamefont {Wang}, \citenamefont
  {Yin}, \citenamefont {Jiang}, \citenamefont {Liu}, \citenamefont {Chen},
  \citenamefont {Huang},\ and\ \citenamefont
  {Duan}}]{cheng2014electroluminescence}%
  \BibitemOpen
  \bibfield  {author} {\bibinfo {author} {\bibfnamefont {R.}~\bibnamefont
  {Cheng}}, \bibinfo {author} {\bibfnamefont {D.}~\bibnamefont {Li}}, \bibinfo
  {author} {\bibfnamefont {H.}~\bibnamefont {Zhou}}, \bibinfo {author}
  {\bibfnamefont {C.}~\bibnamefont {Wang}}, \bibinfo {author} {\bibfnamefont
  {A.}~\bibnamefont {Yin}}, \bibinfo {author} {\bibfnamefont {S.}~\bibnamefont
  {Jiang}}, \bibinfo {author} {\bibfnamefont {Y.}~\bibnamefont {Liu}}, \bibinfo
  {author} {\bibfnamefont {Y.}~\bibnamefont {Chen}}, \bibinfo {author}
  {\bibfnamefont {Y.}~\bibnamefont {Huang}},\ and\ \bibinfo {author}
  {\bibfnamefont {X.}~\bibnamefont {Duan}},\ }\href@noop {} {\bibfield
  {journal} {\bibinfo  {journal} {Nano letters}\ }\textbf {\bibinfo {volume}
  {14}},\ \bibinfo {pages} {5590} (\bibinfo {year} {2014})}\BibitemShut
  {NoStop}%
\bibitem [{\citenamefont {Li}\ \emph {et~al.}(2015)\citenamefont {Li},
  \citenamefont {Shi}, \citenamefont {Cheng}, \citenamefont {Lu}, \citenamefont
  {Lin}, \citenamefont {Tang}, \citenamefont {Tsai}, \citenamefont {Chu},
  \citenamefont {Wei}, \citenamefont {He} \emph {et~al.}}]{li2015epitaxial}%
  \BibitemOpen
  \bibfield  {author} {\bibinfo {author} {\bibfnamefont {M.-Y.}\ \bibnamefont
  {Li}}, \bibinfo {author} {\bibfnamefont {Y.}~\bibnamefont {Shi}}, \bibinfo
  {author} {\bibfnamefont {C.-C.}\ \bibnamefont {Cheng}}, \bibinfo {author}
  {\bibfnamefont {L.-S.}\ \bibnamefont {Lu}}, \bibinfo {author} {\bibfnamefont
  {Y.-C.}\ \bibnamefont {Lin}}, \bibinfo {author} {\bibfnamefont {H.-L.}\
  \bibnamefont {Tang}}, \bibinfo {author} {\bibfnamefont {M.-L.}\ \bibnamefont
  {Tsai}}, \bibinfo {author} {\bibfnamefont {C.-W.}\ \bibnamefont {Chu}},
  \bibinfo {author} {\bibfnamefont {K.-H.}\ \bibnamefont {Wei}}, \bibinfo
  {author} {\bibfnamefont {J.-H.}\ \bibnamefont {He}}, \emph {et~al.},\
  }\href@noop {} {\bibfield  {journal} {\bibinfo  {journal} {Science}\ }\textbf
  {\bibinfo {volume} {349}},\ \bibinfo {pages} {524} (\bibinfo {year}
  {2015})}\BibitemShut {NoStop}%
\bibitem [{\citenamefont {Duong}\ \emph {et~al.}(2019)\citenamefont {Duong},
  \citenamefont {Lee}, \citenamefont {Bang}, \citenamefont {Park},
  \citenamefont {Lim},\ and\ \citenamefont {Jeong}}]{duong2019modulating}%
  \BibitemOpen
  \bibfield  {author} {\bibinfo {author} {\bibfnamefont {N.~T.}\ \bibnamefont
  {Duong}}, \bibinfo {author} {\bibfnamefont {J.}~\bibnamefont {Lee}}, \bibinfo
  {author} {\bibfnamefont {S.}~\bibnamefont {Bang}}, \bibinfo {author}
  {\bibfnamefont {C.}~\bibnamefont {Park}}, \bibinfo {author} {\bibfnamefont
  {S.~C.}\ \bibnamefont {Lim}},\ and\ \bibinfo {author} {\bibfnamefont {M.~S.}\
  \bibnamefont {Jeong}},\ }\href@noop {} {\bibfield  {journal} {\bibinfo
  {journal} {ACS nano}\ }\textbf {\bibinfo {volume} {13}},\ \bibinfo {pages}
  {4478} (\bibinfo {year} {2019})}\BibitemShut {NoStop}%
\bibitem [{\citenamefont {Paul}\ \emph {et~al.}(2017)\citenamefont {Paul},
  \citenamefont {Kuiri}, \citenamefont {Saha}, \citenamefont {Chakraborty},
  \citenamefont {Mahapatra}, \citenamefont {Sood},\ and\ \citenamefont
  {Das}}]{paul2017photo}%
  \BibitemOpen
  \bibfield  {author} {\bibinfo {author} {\bibfnamefont {A.~K.}\ \bibnamefont
  {Paul}}, \bibinfo {author} {\bibfnamefont {M.}~\bibnamefont {Kuiri}},
  \bibinfo {author} {\bibfnamefont {D.}~\bibnamefont {Saha}}, \bibinfo {author}
  {\bibfnamefont {B.}~\bibnamefont {Chakraborty}}, \bibinfo {author}
  {\bibfnamefont {S.}~\bibnamefont {Mahapatra}}, \bibinfo {author}
  {\bibfnamefont {A.}~\bibnamefont {Sood}},\ and\ \bibinfo {author}
  {\bibfnamefont {A.}~\bibnamefont {Das}},\ }\href@noop {} {\bibfield
  {journal} {\bibinfo  {journal} {npj 2D Materials and Applications}\ }\textbf
  {\bibinfo {volume} {1}},\ \bibinfo {pages} {17} (\bibinfo {year}
  {2017})}\BibitemShut {NoStop}%
\bibitem [{\citenamefont {Meng}\ \emph {et~al.}(2024)\citenamefont {Meng},
  \citenamefont {Wang}, \citenamefont {Wang}, \citenamefont {Li}, \citenamefont
  {Zhang},\ and\ \citenamefont {Ho}}]{meng2024anti}%
  \BibitemOpen
  \bibfield  {author} {\bibinfo {author} {\bibfnamefont {Y.}~\bibnamefont
  {Meng}}, \bibinfo {author} {\bibfnamefont {W.}~\bibnamefont {Wang}}, \bibinfo
  {author} {\bibfnamefont {W.}~\bibnamefont {Wang}}, \bibinfo {author}
  {\bibfnamefont {B.}~\bibnamefont {Li}}, \bibinfo {author} {\bibfnamefont
  {Y.}~\bibnamefont {Zhang}},\ and\ \bibinfo {author} {\bibfnamefont
  {J.}~\bibnamefont {Ho}},\ }\href@noop {} {\bibfield  {journal} {\bibinfo
  {journal} {Advanced Materials}\ }\textbf {\bibinfo {volume} {36}},\ \bibinfo
  {pages} {2306290} (\bibinfo {year} {2024})}\BibitemShut {NoStop}%
\bibitem [{\citenamefont {Jadwiszczak}\ \emph {et~al.}(2022)\citenamefont
  {Jadwiszczak}, \citenamefont {Sherman}, \citenamefont {Lynall}, \citenamefont
  {Liu}, \citenamefont {Penkov}, \citenamefont {Young}, \citenamefont
  {Keneipp}, \citenamefont {Drndi\'{c}}, \citenamefont {Hone},\ and\
  \citenamefont {Shepard}}]{jadwiszczak2022mixed}%
  \BibitemOpen
  \bibfield  {author} {\bibinfo {author} {\bibfnamefont {J.}~\bibnamefont
  {Jadwiszczak}}, \bibinfo {author} {\bibfnamefont {J.}~\bibnamefont
  {Sherman}}, \bibinfo {author} {\bibfnamefont {D.}~\bibnamefont {Lynall}},
  \bibinfo {author} {\bibfnamefont {Y.}~\bibnamefont {Liu}}, \bibinfo {author}
  {\bibfnamefont {B.}~\bibnamefont {Penkov}}, \bibinfo {author} {\bibfnamefont
  {E.}~\bibnamefont {Young}}, \bibinfo {author} {\bibfnamefont
  {R.}~\bibnamefont {Keneipp}}, \bibinfo {author} {\bibfnamefont
  {M.}~\bibnamefont {Drndi\'{c}}}, \bibinfo {author} {\bibfnamefont {J.~C.}\
  \bibnamefont {Hone}},\ and\ \bibinfo {author} {\bibfnamefont {K.~L.}\
  \bibnamefont {Shepard}},\ }\href@noop {} {\bibfield  {journal} {\bibinfo
  {journal} {ACS nano}\ }\textbf {\bibinfo {volume} {16}},\ \bibinfo {pages}
  {1639} (\bibinfo {year} {2022})}\BibitemShut {NoStop}%
\bibitem [{\citenamefont {Lu}\ \emph {et~al.}(2021)\citenamefont {Lu},
  \citenamefont {Wei}, \citenamefont {Li}, \citenamefont {Zhang}, \citenamefont
  {Wang}, \citenamefont {Liang}, \citenamefont {Li}, \citenamefont {Fan},\ and\
  \citenamefont {Zhang}}]{lu2021reconfigurable}%
  \BibitemOpen
  \bibfield  {author} {\bibinfo {author} {\bibfnamefont {G.}~\bibnamefont
  {Lu}}, \bibinfo {author} {\bibfnamefont {Y.}~\bibnamefont {Wei}}, \bibinfo
  {author} {\bibfnamefont {X.}~\bibnamefont {Li}}, \bibinfo {author}
  {\bibfnamefont {G.}~\bibnamefont {Zhang}}, \bibinfo {author} {\bibfnamefont
  {G.}~\bibnamefont {Wang}}, \bibinfo {author} {\bibfnamefont {L.}~\bibnamefont
  {Liang}}, \bibinfo {author} {\bibfnamefont {Q.}~\bibnamefont {Li}}, \bibinfo
  {author} {\bibfnamefont {S.}~\bibnamefont {Fan}},\ and\ \bibinfo {author}
  {\bibfnamefont {Y.}~\bibnamefont {Zhang}},\ }\href@noop {} {\bibfield
  {journal} {\bibinfo  {journal} {Nano Letters}\ }\textbf {\bibinfo {volume}
  {21}},\ \bibinfo {pages} {6843} (\bibinfo {year} {2021})}\BibitemShut
  {NoStop}%
\bibitem [{\citenamefont {Wang}\ \emph
  {et~al.}(2022{\natexlab{a}})\citenamefont {Wang}, \citenamefont {Wang},
  \citenamefont {Meng}, \citenamefont {Quan}, \citenamefont {Lai},
  \citenamefont {Li}, \citenamefont {Xie}, \citenamefont {Yip}, \citenamefont
  {Kang}, \citenamefont {Bu} \emph {et~al.}}]{wang2022mixed}%
  \BibitemOpen
  \bibfield  {author} {\bibinfo {author} {\bibfnamefont {W.}~\bibnamefont
  {Wang}}, \bibinfo {author} {\bibfnamefont {W.}~\bibnamefont {Wang}}, \bibinfo
  {author} {\bibfnamefont {Y.}~\bibnamefont {Meng}}, \bibinfo {author}
  {\bibfnamefont {Q.}~\bibnamefont {Quan}}, \bibinfo {author} {\bibfnamefont
  {Z.}~\bibnamefont {Lai}}, \bibinfo {author} {\bibfnamefont {D.}~\bibnamefont
  {Li}}, \bibinfo {author} {\bibfnamefont {P.}~\bibnamefont {Xie}}, \bibinfo
  {author} {\bibfnamefont {S.}~\bibnamefont {Yip}}, \bibinfo {author}
  {\bibfnamefont {X.}~\bibnamefont {Kang}}, \bibinfo {author} {\bibfnamefont
  {X.}~\bibnamefont {Bu}}, \emph {et~al.},\ }\href@noop {} {\bibfield
  {journal} {\bibinfo  {journal} {ACS nano}\ }\textbf {\bibinfo {volume}
  {16}},\ \bibinfo {pages} {11036} (\bibinfo {year}
  {2022}{\natexlab{a}})}\BibitemShut {NoStop}%
\bibitem [{\citenamefont {Wang}\ \emph
  {et~al.}(2022{\natexlab{b}})\citenamefont {Wang}, \citenamefont {Meng},
  \citenamefont {Wang}, \citenamefont {Zhang}, \citenamefont {Xie},
  \citenamefont {Lai}, \citenamefont {Bu}, \citenamefont {Li}, \citenamefont
  {Liu}, \citenamefont {Yang} \emph {et~al.}}]{wang2022highly}%
  \BibitemOpen
  \bibfield  {author} {\bibinfo {author} {\bibfnamefont {W.}~\bibnamefont
  {Wang}}, \bibinfo {author} {\bibfnamefont {Y.}~\bibnamefont {Meng}}, \bibinfo
  {author} {\bibfnamefont {W.}~\bibnamefont {Wang}}, \bibinfo {author}
  {\bibfnamefont {Z.}~\bibnamefont {Zhang}}, \bibinfo {author} {\bibfnamefont
  {P.}~\bibnamefont {Xie}}, \bibinfo {author} {\bibfnamefont {Z.}~\bibnamefont
  {Lai}}, \bibinfo {author} {\bibfnamefont {X.}~\bibnamefont {Bu}}, \bibinfo
  {author} {\bibfnamefont {Y.}~\bibnamefont {Li}}, \bibinfo {author}
  {\bibfnamefont {C.}~\bibnamefont {Liu}}, \bibinfo {author} {\bibfnamefont
  {Z.}~\bibnamefont {Yang}}, \emph {et~al.},\ }\href@noop {} {\bibfield
  {journal} {\bibinfo  {journal} {Advanced Functional Materials}\ }\textbf
  {\bibinfo {volume} {32}},\ \bibinfo {pages} {2203003} (\bibinfo {year}
  {2022}{\natexlab{b}})}\BibitemShut {NoStop}%
\bibitem [{\citenamefont {Zhou}\ \emph {et~al.}(2018)\citenamefont {Zhou},
  \citenamefont {Shu}, \citenamefont {Hu}, \citenamefont {Jiang}, \citenamefont
  {Liang},\ and\ \citenamefont {Chen}}]{zhou2018unveiling}%
  \BibitemOpen
  \bibfield  {author} {\bibinfo {author} {\bibfnamefont {D.}~\bibnamefont
  {Zhou}}, \bibinfo {author} {\bibfnamefont {H.}~\bibnamefont {Shu}}, \bibinfo
  {author} {\bibfnamefont {C.}~\bibnamefont {Hu}}, \bibinfo {author}
  {\bibfnamefont {L.}~\bibnamefont {Jiang}}, \bibinfo {author} {\bibfnamefont
  {P.}~\bibnamefont {Liang}},\ and\ \bibinfo {author} {\bibfnamefont
  {X.}~\bibnamefont {Chen}},\ }\href@noop {} {\bibfield  {journal} {\bibinfo
  {journal} {Crystal Growth \& Design}\ }\textbf {\bibinfo {volume} {18}},\
  \bibinfo {pages} {1012} (\bibinfo {year} {2018})}\BibitemShut {NoStop}%
\bibitem [{\citenamefont {Wang}\ \emph {et~al.}(2014)\citenamefont {Wang},
  \citenamefont {Safdar}, \citenamefont {Xu}, \citenamefont {Mirza},
  \citenamefont {Wang},\ and\ \citenamefont {He}}]{wang2014van}%
  \BibitemOpen
  \bibfield  {author} {\bibinfo {author} {\bibfnamefont {Q.}~\bibnamefont
  {Wang}}, \bibinfo {author} {\bibfnamefont {M.}~\bibnamefont {Safdar}},
  \bibinfo {author} {\bibfnamefont {K.}~\bibnamefont {Xu}}, \bibinfo {author}
  {\bibfnamefont {M.}~\bibnamefont {Mirza}}, \bibinfo {author} {\bibfnamefont
  {Z.}~\bibnamefont {Wang}},\ and\ \bibinfo {author} {\bibfnamefont
  {J.}~\bibnamefont {He}},\ }\href@noop {} {\bibfield  {journal} {\bibinfo
  {journal} {ACS nano}\ }\textbf {\bibinfo {volume} {8}},\ \bibinfo {pages}
  {7497} (\bibinfo {year} {2014})}\BibitemShut {NoStop}%
\bibitem [{\citenamefont {Pine}\ and\ \citenamefont
  {Dresselhaus}(1971)}]{pine1971raman}%
  \BibitemOpen
  \bibfield  {author} {\bibinfo {author} {\bibfnamefont {A.}~\bibnamefont
  {Pine}}\ and\ \bibinfo {author} {\bibfnamefont {G.}~\bibnamefont
  {Dresselhaus}},\ }\href@noop {} {\bibfield  {journal} {\bibinfo  {journal}
  {Physical Review B}\ }\textbf {\bibinfo {volume} {4}},\ \bibinfo {pages}
  {356} (\bibinfo {year} {1971})}\BibitemShut {NoStop}%
\bibitem [{\citenamefont {Qin}\ \emph {et~al.}(2020)\citenamefont {Qin},
  \citenamefont {Liao}, \citenamefont {Si}, \citenamefont {Gao}, \citenamefont
  {Qiu}, \citenamefont {Jian}, \citenamefont {Wang}, \citenamefont {Zhang},
  \citenamefont {Huang}, \citenamefont {Charnas} \emph
  {et~al.}}]{qin2020raman}%
  \BibitemOpen
  \bibfield  {author} {\bibinfo {author} {\bibfnamefont {J.-K.}\ \bibnamefont
  {Qin}}, \bibinfo {author} {\bibfnamefont {P.-Y.}\ \bibnamefont {Liao}},
  \bibinfo {author} {\bibfnamefont {M.}~\bibnamefont {Si}}, \bibinfo {author}
  {\bibfnamefont {S.}~\bibnamefont {Gao}}, \bibinfo {author} {\bibfnamefont
  {G.}~\bibnamefont {Qiu}}, \bibinfo {author} {\bibfnamefont {J.}~\bibnamefont
  {Jian}}, \bibinfo {author} {\bibfnamefont {Q.}~\bibnamefont {Wang}}, \bibinfo
  {author} {\bibfnamefont {S.-Q.}\ \bibnamefont {Zhang}}, \bibinfo {author}
  {\bibfnamefont {S.}~\bibnamefont {Huang}}, \bibinfo {author} {\bibfnamefont
  {A.}~\bibnamefont {Charnas}}, \emph {et~al.},\ }\href@noop {} {\bibfield
  {journal} {\bibinfo  {journal} {Nature electronics}\ }\textbf {\bibinfo
  {volume} {3}},\ \bibinfo {pages} {141} (\bibinfo {year} {2020})}\BibitemShut
  {NoStop}%
\bibitem [{\citenamefont {Du}\ \emph {et~al.}(2017)\citenamefont {Du},
  \citenamefont {Qiu}, \citenamefont {Wang}, \citenamefont {Si}, \citenamefont
  {Xu}, \citenamefont {Wu},\ and\ \citenamefont {Ye}}]{du2017one}%
  \BibitemOpen
  \bibfield  {author} {\bibinfo {author} {\bibfnamefont {Y.}~\bibnamefont
  {Du}}, \bibinfo {author} {\bibfnamefont {G.}~\bibnamefont {Qiu}}, \bibinfo
  {author} {\bibfnamefont {Y.}~\bibnamefont {Wang}}, \bibinfo {author}
  {\bibfnamefont {M.}~\bibnamefont {Si}}, \bibinfo {author} {\bibfnamefont
  {X.}~\bibnamefont {Xu}}, \bibinfo {author} {\bibfnamefont {W.}~\bibnamefont
  {Wu}},\ and\ \bibinfo {author} {\bibfnamefont {P.~D.}\ \bibnamefont {Ye}},\
  }\href@noop {} {\bibfield  {journal} {\bibinfo  {journal} {Nano letters}\
  }\textbf {\bibinfo {volume} {17}},\ \bibinfo {pages} {3965} (\bibinfo {year}
  {2017})}\BibitemShut {NoStop}%
\bibitem [{\citenamefont {Maguire}\ \emph {et~al.}(2019)\citenamefont
  {Maguire}, \citenamefont {Jadwiszczak}, \citenamefont {O’Brien},
  \citenamefont {Keane}, \citenamefont {Duesberg}, \citenamefont {McEvoy},\
  and\ \citenamefont {Zhang}}]{maguire2019defect}%
  \BibitemOpen
  \bibfield  {author} {\bibinfo {author} {\bibfnamefont {P.}~\bibnamefont
  {Maguire}}, \bibinfo {author} {\bibfnamefont {J.}~\bibnamefont
  {Jadwiszczak}}, \bibinfo {author} {\bibfnamefont {M.}~\bibnamefont
  {O’Brien}}, \bibinfo {author} {\bibfnamefont {D.}~\bibnamefont {Keane}},
  \bibinfo {author} {\bibfnamefont {G.~S.}\ \bibnamefont {Duesberg}}, \bibinfo
  {author} {\bibfnamefont {N.}~\bibnamefont {McEvoy}},\ and\ \bibinfo {author}
  {\bibfnamefont {H.}~\bibnamefont {Zhang}},\ }\href@noop {} {\bibfield
  {journal} {\bibinfo  {journal} {Journal of Applied Physics}\ }\textbf
  {\bibinfo {volume} {126}} (\bibinfo {year} {2019})}\BibitemShut {NoStop}%
\bibitem [{\citenamefont {Lee}\ \emph {et~al.}(2009)\citenamefont {Lee},
  \citenamefont {Yoo}, \citenamefont {Choi}, \citenamefont {Kang},
  \citenamefont {Jeon}, \citenamefont {Kim}, \citenamefont {Seo},\ and\
  \citenamefont {Chung}}]{lee2009interlayer}%
  \BibitemOpen
  \bibfield  {author} {\bibinfo {author} {\bibfnamefont {N.}~\bibnamefont
  {Lee}}, \bibinfo {author} {\bibfnamefont {J.}~\bibnamefont {Yoo}}, \bibinfo
  {author} {\bibfnamefont {Y.}~\bibnamefont {Choi}}, \bibinfo {author}
  {\bibfnamefont {C.}~\bibnamefont {Kang}}, \bibinfo {author} {\bibfnamefont
  {D.}~\bibnamefont {Jeon}}, \bibinfo {author} {\bibfnamefont {D.}~\bibnamefont
  {Kim}}, \bibinfo {author} {\bibfnamefont {S.}~\bibnamefont {Seo}},\ and\
  \bibinfo {author} {\bibfnamefont {H.}~\bibnamefont {Chung}},\ }\href@noop {}
  {\bibfield  {journal} {\bibinfo  {journal} {Applied Physics Letters}\
  }\textbf {\bibinfo {volume} {95}} (\bibinfo {year} {2009})}\BibitemShut
  {NoStop}%
\bibitem [{\citenamefont {Choi}\ \emph {et~al.}(2014)\citenamefont {Choi},
  \citenamefont {Shaolin},\ and\ \citenamefont {Yang}}]{choi2014layer}%
  \BibitemOpen
  \bibfield  {author} {\bibinfo {author} {\bibfnamefont {S.}~\bibnamefont
  {Choi}}, \bibinfo {author} {\bibfnamefont {Z.}~\bibnamefont {Shaolin}},\ and\
  \bibinfo {author} {\bibfnamefont {W.}~\bibnamefont {Yang}},\ }\href@noop {}
  {\bibfield  {journal} {\bibinfo  {journal} {Journal of the Korean Physical
  Society}\ }\textbf {\bibinfo {volume} {64}},\ \bibinfo {pages} {1550}
  (\bibinfo {year} {2014})}\BibitemShut {NoStop}%
\bibitem [{\citenamefont {Michaelson}(1977)}]{michaelson1977work}%
  \BibitemOpen
  \bibfield  {author} {\bibinfo {author} {\bibfnamefont {H.~B.}\ \bibnamefont
  {Michaelson}},\ }\href@noop {} {\bibfield  {journal} {\bibinfo  {journal}
  {Journal of applied physics}\ }\textbf {\bibinfo {volume} {48}},\ \bibinfo
  {pages} {4729} (\bibinfo {year} {1977})}\BibitemShut {NoStop}%
\bibitem [{\citenamefont {Ford}\ \emph {et~al.}(2009)\citenamefont {Ford},
  \citenamefont {Ho}, \citenamefont {Chueh}, \citenamefont {Tseng},
  \citenamefont {Fan}, \citenamefont {Guo}, \citenamefont {Bokor},\ and\
  \citenamefont {Javey}}]{ford2009diameter}%
  \BibitemOpen
  \bibfield  {author} {\bibinfo {author} {\bibfnamefont {A.~C.}\ \bibnamefont
  {Ford}}, \bibinfo {author} {\bibfnamefont {J.~C.}\ \bibnamefont {Ho}},
  \bibinfo {author} {\bibfnamefont {Y.-L.}\ \bibnamefont {Chueh}}, \bibinfo
  {author} {\bibfnamefont {Y.-C.}\ \bibnamefont {Tseng}}, \bibinfo {author}
  {\bibfnamefont {Z.}~\bibnamefont {Fan}}, \bibinfo {author} {\bibfnamefont
  {J.}~\bibnamefont {Guo}}, \bibinfo {author} {\bibfnamefont {J.}~\bibnamefont
  {Bokor}},\ and\ \bibinfo {author} {\bibfnamefont {A.}~\bibnamefont {Javey}},\
  }\href@noop {} {\bibfield  {journal} {\bibinfo  {journal} {Nano Letters}\
  }\textbf {\bibinfo {volume} {9}},\ \bibinfo {pages} {360} (\bibinfo {year}
  {2009})}\BibitemShut {NoStop}%
\bibitem [{\citenamefont {Sharma}\ \emph {et~al.}(2022)\citenamefont {Sharma},
  \citenamefont {Thakur},\ and\ \citenamefont {Sharma}}]{sharma2022structural}%
  \BibitemOpen
  \bibfield  {author} {\bibinfo {author} {\bibfnamefont {T.}~\bibnamefont
  {Sharma}}, \bibinfo {author} {\bibfnamefont {R.}~\bibnamefont {Thakur}},\
  and\ \bibinfo {author} {\bibfnamefont {R.}~\bibnamefont {Sharma}},\
  }\href@noop {} {\bibfield  {journal} {\bibinfo  {journal} {Applied Physics
  A}\ }\textbf {\bibinfo {volume} {128}},\ \bibinfo {pages} {1} (\bibinfo
  {year} {2022})}\BibitemShut {NoStop}%
\bibitem [{\citenamefont {Santos}\ and\ \citenamefont
  {Kaxiras}(2013)}]{santos2013electrically}%
  \BibitemOpen
  \bibfield  {author} {\bibinfo {author} {\bibfnamefont {E.~J.}\ \bibnamefont
  {Santos}}\ and\ \bibinfo {author} {\bibfnamefont {E.}~\bibnamefont
  {Kaxiras}},\ }\href@noop {} {\bibfield  {journal} {\bibinfo  {journal} {ACS
  nano}\ }\textbf {\bibinfo {volume} {7}},\ \bibinfo {pages} {10741} (\bibinfo
  {year} {2013})}\BibitemShut {NoStop}%
\bibitem [{\citenamefont {Chu}\ \emph {et~al.}(2018)\citenamefont {Chu},
  \citenamefont {Wang}, \citenamefont {Lee}, \citenamefont {Chen},
  \citenamefont {Yang}, \citenamefont {Butler}, \citenamefont {Lu},
  \citenamefont {Yeh}, \citenamefont {Chang},\ and\ \citenamefont
  {Lin}}]{chu2018atomic}%
  \BibitemOpen
  \bibfield  {author} {\bibinfo {author} {\bibfnamefont {Y.-H.}\ \bibnamefont
  {Chu}}, \bibinfo {author} {\bibfnamefont {L.-H.}\ \bibnamefont {Wang}},
  \bibinfo {author} {\bibfnamefont {S.-Y.}\ \bibnamefont {Lee}}, \bibinfo
  {author} {\bibfnamefont {H.-J.}\ \bibnamefont {Chen}}, \bibinfo {author}
  {\bibfnamefont {P.-Y.}\ \bibnamefont {Yang}}, \bibinfo {author}
  {\bibfnamefont {C.~J.}\ \bibnamefont {Butler}}, \bibinfo {author}
  {\bibfnamefont {L.-S.}\ \bibnamefont {Lu}}, \bibinfo {author} {\bibfnamefont
  {H.}~\bibnamefont {Yeh}}, \bibinfo {author} {\bibfnamefont {W.-H.}\
  \bibnamefont {Chang}},\ and\ \bibinfo {author} {\bibfnamefont {M.-T.}\
  \bibnamefont {Lin}},\ }\href@noop {} {\bibfield  {journal} {\bibinfo
  {journal} {Applied Physics Letters}\ }\textbf {\bibinfo {volume} {113}}
  (\bibinfo {year} {2018})}\BibitemShut {NoStop}%
\bibitem [{\citenamefont {Yao}\ \emph {et~al.}(2021)\citenamefont {Yao},
  \citenamefont {Chen}, \citenamefont {Li}, \citenamefont {Du}, \citenamefont
  {Wu}, \citenamefont {Tian}, \citenamefont {Zhang}, \citenamefont {Yang},
  \citenamefont {Li},\ and\ \citenamefont {Lin}}]{yao2021high}%
  \BibitemOpen
  \bibfield  {author} {\bibinfo {author} {\bibfnamefont {J.}~\bibnamefont
  {Yao}}, \bibinfo {author} {\bibfnamefont {F.}~\bibnamefont {Chen}}, \bibinfo
  {author} {\bibfnamefont {J.}~\bibnamefont {Li}}, \bibinfo {author}
  {\bibfnamefont {J.}~\bibnamefont {Du}}, \bibinfo {author} {\bibfnamefont
  {D.}~\bibnamefont {Wu}}, \bibinfo {author} {\bibfnamefont {Y.}~\bibnamefont
  {Tian}}, \bibinfo {author} {\bibfnamefont {C.}~\bibnamefont {Zhang}},
  \bibinfo {author} {\bibfnamefont {J.}~\bibnamefont {Yang}}, \bibinfo {author}
  {\bibfnamefont {X.}~\bibnamefont {Li}},\ and\ \bibinfo {author}
  {\bibfnamefont {P.}~\bibnamefont {Lin}},\ }\href@noop {} {\bibfield
  {journal} {\bibinfo  {journal} {Journal of Materials Chemistry C}\ }\textbf
  {\bibinfo {volume} {9}},\ \bibinfo {pages} {13123} (\bibinfo {year}
  {2021})}\BibitemShut {NoStop}%
\bibitem [{\citenamefont {Huang}\ \emph {et~al.}(2018)\citenamefont {Huang},
  \citenamefont {Zhuge}, \citenamefont {Hou}, \citenamefont {Lv}, \citenamefont
  {Luo}, \citenamefont {Zhou}, \citenamefont {Gan},\ and\ \citenamefont
  {Zhai}}]{huang2018van}%
  \BibitemOpen
  \bibfield  {author} {\bibinfo {author} {\bibfnamefont {Y.}~\bibnamefont
  {Huang}}, \bibinfo {author} {\bibfnamefont {F.}~\bibnamefont {Zhuge}},
  \bibinfo {author} {\bibfnamefont {J.}~\bibnamefont {Hou}}, \bibinfo {author}
  {\bibfnamefont {L.}~\bibnamefont {Lv}}, \bibinfo {author} {\bibfnamefont
  {P.}~\bibnamefont {Luo}}, \bibinfo {author} {\bibfnamefont {N.}~\bibnamefont
  {Zhou}}, \bibinfo {author} {\bibfnamefont {L.}~\bibnamefont {Gan}},\ and\
  \bibinfo {author} {\bibfnamefont {T.}~\bibnamefont {Zhai}},\ }\href@noop {}
  {\bibfield  {journal} {\bibinfo  {journal} {ACS nano}\ }\textbf {\bibinfo
  {volume} {12}},\ \bibinfo {pages} {4062} (\bibinfo {year}
  {2018})}\BibitemShut {NoStop}%
\bibitem [{\citenamefont {Stradi}\ \emph {et~al.}(2016)\citenamefont {Stradi},
  \citenamefont {Martinez}, \citenamefont {Blom}, \citenamefont {Brandbyge},\
  and\ \citenamefont {Stokbro}}]{stradi2016general}%
  \BibitemOpen
  \bibfield  {author} {\bibinfo {author} {\bibfnamefont {D.}~\bibnamefont
  {Stradi}}, \bibinfo {author} {\bibfnamefont {U.}~\bibnamefont {Martinez}},
  \bibinfo {author} {\bibfnamefont {A.}~\bibnamefont {Blom}}, \bibinfo {author}
  {\bibfnamefont {M.}~\bibnamefont {Brandbyge}},\ and\ \bibinfo {author}
  {\bibfnamefont {K.}~\bibnamefont {Stokbro}},\ }\href@noop {} {\bibfield
  {journal} {\bibinfo  {journal} {Physical Review B}\ }\textbf {\bibinfo
  {volume} {93}},\ \bibinfo {pages} {155302} (\bibinfo {year}
  {2016})}\BibitemShut {NoStop}%
\bibitem [{\citenamefont {Soler}\ \emph {et~al.}(2002)\citenamefont {Soler},
  \citenamefont {Artacho}, \citenamefont {Gale}, \citenamefont {Garc{\'\i}a},
  \citenamefont {Junquera}, \citenamefont {Ordej{\'o}n},\ and\ \citenamefont
  {S{\'a}nchez-Portal}}]{soler2002siesta}%
  \BibitemOpen
  \bibfield  {author} {\bibinfo {author} {\bibfnamefont {J.~M.}\ \bibnamefont
  {Soler}}, \bibinfo {author} {\bibfnamefont {E.}~\bibnamefont {Artacho}},
  \bibinfo {author} {\bibfnamefont {J.~D.}\ \bibnamefont {Gale}}, \bibinfo
  {author} {\bibfnamefont {A.}~\bibnamefont {Garc{\'\i}a}}, \bibinfo {author}
  {\bibfnamefont {J.}~\bibnamefont {Junquera}}, \bibinfo {author}
  {\bibfnamefont {P.}~\bibnamefont {Ordej{\'o}n}},\ and\ \bibinfo {author}
  {\bibfnamefont {D.}~\bibnamefont {S{\'a}nchez-Portal}},\ }\href@noop {}
  {\bibfield  {journal} {\bibinfo  {journal} {Journal of Physics: Condensed
  Matter}\ }\textbf {\bibinfo {volume} {14}},\ \bibinfo {pages} {2745}
  (\bibinfo {year} {2002})}\BibitemShut {NoStop}%
\bibitem [{\citenamefont {Wang}\ \emph {et~al.}(2015)\citenamefont {Wang},
  \citenamefont {Stepanov}, \citenamefont {Gray}, \citenamefont {Lau},
  \citenamefont {Itkis},\ and\ \citenamefont {Haddon}}]{wang2015ionic}%
  \BibitemOpen
  \bibfield  {author} {\bibinfo {author} {\bibfnamefont {F.}~\bibnamefont
  {Wang}}, \bibinfo {author} {\bibfnamefont {P.}~\bibnamefont {Stepanov}},
  \bibinfo {author} {\bibfnamefont {M.}~\bibnamefont {Gray}}, \bibinfo {author}
  {\bibfnamefont {C.~N.}\ \bibnamefont {Lau}}, \bibinfo {author} {\bibfnamefont
  {M.~E.}\ \bibnamefont {Itkis}},\ and\ \bibinfo {author} {\bibfnamefont
  {R.~C.}\ \bibnamefont {Haddon}},\ }\href@noop {} {\bibfield  {journal}
  {\bibinfo  {journal} {Nano letters}\ }\textbf {\bibinfo {volume} {15}},\
  \bibinfo {pages} {5284} (\bibinfo {year} {2015})}\BibitemShut {NoStop}%
\bibitem [{\citenamefont {Lieb}\ \emph {et~al.}(2019)\citenamefont {Lieb},
  \citenamefont {Demontis}, \citenamefont {Prete}, \citenamefont {Ercolani},
  \citenamefont {Zannier}, \citenamefont {Sorba}, \citenamefont {Ono},
  \citenamefont {Beltram}, \citenamefont {Sac{\'e}p{\'e}},\ and\ \citenamefont
  {Rossella}}]{lieb2019ionic}%
  \BibitemOpen
  \bibfield  {author} {\bibinfo {author} {\bibfnamefont {J.}~\bibnamefont
  {Lieb}}, \bibinfo {author} {\bibfnamefont {V.}~\bibnamefont {Demontis}},
  \bibinfo {author} {\bibfnamefont {D.}~\bibnamefont {Prete}}, \bibinfo
  {author} {\bibfnamefont {D.}~\bibnamefont {Ercolani}}, \bibinfo {author}
  {\bibfnamefont {V.}~\bibnamefont {Zannier}}, \bibinfo {author} {\bibfnamefont
  {L.}~\bibnamefont {Sorba}}, \bibinfo {author} {\bibfnamefont
  {S.}~\bibnamefont {Ono}}, \bibinfo {author} {\bibfnamefont {F.}~\bibnamefont
  {Beltram}}, \bibinfo {author} {\bibfnamefont {B.}~\bibnamefont
  {Sac{\'e}p{\'e}}},\ and\ \bibinfo {author} {\bibfnamefont {F.}~\bibnamefont
  {Rossella}},\ }\href@noop {} {\bibfield  {journal} {\bibinfo  {journal}
  {Advanced Functional Materials}\ }\textbf {\bibinfo {volume} {29}},\ \bibinfo
  {pages} {1804378} (\bibinfo {year} {2019})}\BibitemShut {NoStop}%
\bibitem [{\citenamefont {Jo}\ \emph {et~al.}(2015)\citenamefont {Jo},
  \citenamefont {Costanzo}, \citenamefont {Berger},\ and\ \citenamefont
  {Morpurgo}}]{jo2015electrostatically}%
  \BibitemOpen
  \bibfield  {author} {\bibinfo {author} {\bibfnamefont {S.}~\bibnamefont
  {Jo}}, \bibinfo {author} {\bibfnamefont {D.}~\bibnamefont {Costanzo}},
  \bibinfo {author} {\bibfnamefont {H.}~\bibnamefont {Berger}},\ and\ \bibinfo
  {author} {\bibfnamefont {A.~F.}\ \bibnamefont {Morpurgo}},\ }\href@noop {}
  {\bibfield  {journal} {\bibinfo  {journal} {Nano letters}\ }\textbf {\bibinfo
  {volume} {15}},\ \bibinfo {pages} {1197} (\bibinfo {year}
  {2015})}\BibitemShut {NoStop}%
\bibitem [{\citenamefont {Chien}\ \emph {et~al.}(2016)\citenamefont {Chien},
  \citenamefont {Yuan}, \citenamefont {Wang},\ and\ \citenamefont
  {Lee}}]{chien2016thermoelectric}%
  \BibitemOpen
  \bibfield  {author} {\bibinfo {author} {\bibfnamefont {Y.-Y.}\ \bibnamefont
  {Chien}}, \bibinfo {author} {\bibfnamefont {H.}~\bibnamefont {Yuan}},
  \bibinfo {author} {\bibfnamefont {C.-R.}\ \bibnamefont {Wang}},\ and\
  \bibinfo {author} {\bibfnamefont {W.-L.}\ \bibnamefont {Lee}},\ }\href@noop
  {} {\bibfield  {journal} {\bibinfo  {journal} {Scientific reports}\ }\textbf
  {\bibinfo {volume} {6}},\ \bibinfo {pages} {20402} (\bibinfo {year}
  {2016})}\BibitemShut {NoStop}%
\bibitem [{\citenamefont {Yamada}\ \emph {et~al.}(2011)\citenamefont {Yamada},
  \citenamefont {Ueno}, \citenamefont {Fukumura}, \citenamefont {Yuan},
  \citenamefont {Shimotani}, \citenamefont {Iwasa}, \citenamefont {Gu},
  \citenamefont {Tsukimoto}, \citenamefont {Ikuhara},\ and\ \citenamefont
  {Kawasaki}}]{yamada2011electrically}%
  \BibitemOpen
  \bibfield  {author} {\bibinfo {author} {\bibfnamefont {Y.}~\bibnamefont
  {Yamada}}, \bibinfo {author} {\bibfnamefont {K.}~\bibnamefont {Ueno}},
  \bibinfo {author} {\bibfnamefont {T.}~\bibnamefont {Fukumura}}, \bibinfo
  {author} {\bibfnamefont {H.}~\bibnamefont {Yuan}}, \bibinfo {author}
  {\bibfnamefont {H.}~\bibnamefont {Shimotani}}, \bibinfo {author}
  {\bibfnamefont {Y.}~\bibnamefont {Iwasa}}, \bibinfo {author} {\bibfnamefont
  {L.}~\bibnamefont {Gu}}, \bibinfo {author} {\bibfnamefont {S.}~\bibnamefont
  {Tsukimoto}}, \bibinfo {author} {\bibfnamefont {Y.}~\bibnamefont {Ikuhara}},\
  and\ \bibinfo {author} {\bibfnamefont {M.}~\bibnamefont {Kawasaki}},\
  }\href@noop {} {\bibfield  {journal} {\bibinfo  {journal} {Science}\ }\textbf
  {\bibinfo {volume} {332}},\ \bibinfo {pages} {1065} (\bibinfo {year}
  {2011})}\BibitemShut {NoStop}%
\bibitem [{\citenamefont {Nakano}\ \emph {et~al.}(2012)\citenamefont {Nakano},
  \citenamefont {Shibuya}, \citenamefont {Okuyama}, \citenamefont {Hatano},
  \citenamefont {Ono}, \citenamefont {Kawasaki}, \citenamefont {Iwasa},\ and\
  \citenamefont {Tokura}}]{nakano2012collective}%
  \BibitemOpen
  \bibfield  {author} {\bibinfo {author} {\bibfnamefont {M.}~\bibnamefont
  {Nakano}}, \bibinfo {author} {\bibfnamefont {K.}~\bibnamefont {Shibuya}},
  \bibinfo {author} {\bibfnamefont {D.}~\bibnamefont {Okuyama}}, \bibinfo
  {author} {\bibfnamefont {T.}~\bibnamefont {Hatano}}, \bibinfo {author}
  {\bibfnamefont {S.}~\bibnamefont {Ono}}, \bibinfo {author} {\bibfnamefont
  {M.}~\bibnamefont {Kawasaki}}, \bibinfo {author} {\bibfnamefont
  {Y.}~\bibnamefont {Iwasa}},\ and\ \bibinfo {author} {\bibfnamefont
  {Y.}~\bibnamefont {Tokura}},\ }\href@noop {} {\bibfield  {journal} {\bibinfo
  {journal} {Nature}\ }\textbf {\bibinfo {volume} {487}},\ \bibinfo {pages}
  {459} (\bibinfo {year} {2012})}\BibitemShut {NoStop}%
\bibitem [{\citenamefont {Zhang}\ \emph {et~al.}(2019)\citenamefont {Zhang},
  \citenamefont {Zhang}, \citenamefont {Krylyuk}, \citenamefont {Milligan},
  \citenamefont {Zhu}, \citenamefont {Zemlyanov}, \citenamefont {Bendersky},
  \citenamefont {Burton}, \citenamefont {Davydov},\ and\ \citenamefont
  {Appenzeller}}]{zhang2019electric}%
  \BibitemOpen
  \bibfield  {author} {\bibinfo {author} {\bibfnamefont {F.}~\bibnamefont
  {Zhang}}, \bibinfo {author} {\bibfnamefont {H.}~\bibnamefont {Zhang}},
  \bibinfo {author} {\bibfnamefont {S.}~\bibnamefont {Krylyuk}}, \bibinfo
  {author} {\bibfnamefont {C.~A.}\ \bibnamefont {Milligan}}, \bibinfo {author}
  {\bibfnamefont {Y.}~\bibnamefont {Zhu}}, \bibinfo {author} {\bibfnamefont
  {D.~Y.}\ \bibnamefont {Zemlyanov}}, \bibinfo {author} {\bibfnamefont {L.~A.}\
  \bibnamefont {Bendersky}}, \bibinfo {author} {\bibfnamefont {B.~P.}\
  \bibnamefont {Burton}}, \bibinfo {author} {\bibfnamefont {A.~V.}\
  \bibnamefont {Davydov}},\ and\ \bibinfo {author} {\bibfnamefont
  {J.}~\bibnamefont {Appenzeller}},\ }\href@noop {} {\bibfield  {journal}
  {\bibinfo  {journal} {Nature materials}\ }\textbf {\bibinfo {volume} {18}},\
  \bibinfo {pages} {55} (\bibinfo {year} {2019})}\BibitemShut {NoStop}%
\bibitem [{\citenamefont {Dasika}\ \emph {et~al.}(2021)\citenamefont {Dasika},
  \citenamefont {Samantaray}, \citenamefont {Murali}, \citenamefont {Abraham},
  \citenamefont {Watanbe}, \citenamefont {Taniguchi}, \citenamefont
  {Ravishankar},\ and\ \citenamefont {Majumdar}}]{dasika2021contact}%
  \BibitemOpen
  \bibfield  {author} {\bibinfo {author} {\bibfnamefont {P.}~\bibnamefont
  {Dasika}}, \bibinfo {author} {\bibfnamefont {D.}~\bibnamefont {Samantaray}},
  \bibinfo {author} {\bibfnamefont {K.}~\bibnamefont {Murali}}, \bibinfo
  {author} {\bibfnamefont {N.}~\bibnamefont {Abraham}}, \bibinfo {author}
  {\bibfnamefont {K.}~\bibnamefont {Watanbe}}, \bibinfo {author} {\bibfnamefont
  {T.}~\bibnamefont {Taniguchi}}, \bibinfo {author} {\bibfnamefont
  {N.}~\bibnamefont {Ravishankar}},\ and\ \bibinfo {author} {\bibfnamefont
  {K.}~\bibnamefont {Majumdar}},\ }\href@noop {} {\bibfield  {journal}
  {\bibinfo  {journal} {Advanced Functional Materials}\ }\textbf {\bibinfo
  {volume} {31}},\ \bibinfo {pages} {2006278} (\bibinfo {year}
  {2021})}\BibitemShut {NoStop}%
\bibitem [{\citenamefont {Chu}\ \emph {et~al.}(2014)\citenamefont {Chu},
  \citenamefont {Schmidt}, \citenamefont {Pu}, \citenamefont {Wang},
  \citenamefont {{\"O}zyilmaz}, \citenamefont {Takenobu},\ and\ \citenamefont
  {Eda}}]{chu2014charge}%
  \BibitemOpen
  \bibfield  {author} {\bibinfo {author} {\bibfnamefont {L.}~\bibnamefont
  {Chu}}, \bibinfo {author} {\bibfnamefont {H.}~\bibnamefont {Schmidt}},
  \bibinfo {author} {\bibfnamefont {J.}~\bibnamefont {Pu}}, \bibinfo {author}
  {\bibfnamefont {S.}~\bibnamefont {Wang}}, \bibinfo {author} {\bibfnamefont
  {B.}~\bibnamefont {{\"O}zyilmaz}}, \bibinfo {author} {\bibfnamefont
  {T.}~\bibnamefont {Takenobu}},\ and\ \bibinfo {author} {\bibfnamefont
  {G.}~\bibnamefont {Eda}},\ }\href@noop {} {\bibfield  {journal} {\bibinfo
  {journal} {Scientific reports}\ }\textbf {\bibinfo {volume} {4}},\ \bibinfo
  {pages} {7293} (\bibinfo {year} {2014})}\BibitemShut {NoStop}%
\bibitem [{\citenamefont {Ding}\ \emph {et~al.}(2012)\citenamefont {Ding},
  \citenamefont {Zhang}, \citenamefont {Liang}, \citenamefont {Pei},
  \citenamefont {Wang}, \citenamefont {Li}, \citenamefont {Zhou}, \citenamefont
  {Liu},\ and\ \citenamefont {Peng}}]{ding2012cmos}%
  \BibitemOpen
  \bibfield  {author} {\bibinfo {author} {\bibfnamefont {L.}~\bibnamefont
  {Ding}}, \bibinfo {author} {\bibfnamefont {Z.}~\bibnamefont {Zhang}},
  \bibinfo {author} {\bibfnamefont {S.}~\bibnamefont {Liang}}, \bibinfo
  {author} {\bibfnamefont {T.}~\bibnamefont {Pei}}, \bibinfo {author}
  {\bibfnamefont {S.}~\bibnamefont {Wang}}, \bibinfo {author} {\bibfnamefont
  {Y.}~\bibnamefont {Li}}, \bibinfo {author} {\bibfnamefont {W.}~\bibnamefont
  {Zhou}}, \bibinfo {author} {\bibfnamefont {J.}~\bibnamefont {Liu}},\ and\
  \bibinfo {author} {\bibfnamefont {L.-M.}\ \bibnamefont {Peng}},\ }\href@noop
  {} {\bibfield  {journal} {\bibinfo  {journal} {Nature communications}\
  }\textbf {\bibinfo {volume} {3}},\ \bibinfo {pages} {677} (\bibinfo {year}
  {2012})}\BibitemShut {NoStop}%
\bibitem [{\citenamefont {Kresse}\ and\ \citenamefont
  {Furthm{\"u}ller}(1996{\natexlab{a}})}]{kresse1996efficient}%
  \BibitemOpen
  \bibfield  {author} {\bibinfo {author} {\bibfnamefont {G.}~\bibnamefont
  {Kresse}}\ and\ \bibinfo {author} {\bibfnamefont {J.}~\bibnamefont
  {Furthm{\"u}ller}},\ }\href@noop {} {\bibfield  {journal} {\bibinfo
  {journal} {Physical review B}\ }\textbf {\bibinfo {volume} {54}},\ \bibinfo
  {pages} {11169} (\bibinfo {year} {1996}{\natexlab{a}})}\BibitemShut {NoStop}%
\bibitem [{\citenamefont {Kresse}\ and\ \citenamefont
  {Furthm{\"u}ller}(1996{\natexlab{b}})}]{kresse1996efficiency}%
  \BibitemOpen
  \bibfield  {author} {\bibinfo {author} {\bibfnamefont {G.}~\bibnamefont
  {Kresse}}\ and\ \bibinfo {author} {\bibfnamefont {J.}~\bibnamefont
  {Furthm{\"u}ller}},\ }\href@noop {} {\bibfield  {journal} {\bibinfo
  {journal} {Computational materials science}\ }\textbf {\bibinfo {volume}
  {6}},\ \bibinfo {pages} {15} (\bibinfo {year}
  {1996}{\natexlab{b}})}\BibitemShut {NoStop}%
\bibitem [{\citenamefont {Perdew}\ \emph {et~al.}(1996)\citenamefont {Perdew},
  \citenamefont {Burke},\ and\ \citenamefont
  {Ernzerhof}}]{perdew1996generalized}%
  \BibitemOpen
  \bibfield  {author} {\bibinfo {author} {\bibfnamefont {J.~P.}\ \bibnamefont
  {Perdew}}, \bibinfo {author} {\bibfnamefont {K.}~\bibnamefont {Burke}},\ and\
  \bibinfo {author} {\bibfnamefont {M.}~\bibnamefont {Ernzerhof}},\ }\href@noop
  {} {\bibfield  {journal} {\bibinfo  {journal} {Physical review letters}\
  }\textbf {\bibinfo {volume} {77}},\ \bibinfo {pages} {3865} (\bibinfo {year}
  {1996})}\BibitemShut {NoStop}%
\bibitem [{\citenamefont {{Synopsys QuantumATK}}(2023)}]{QuantumATK2023}%
  \BibitemOpen
  \bibfield  {author} {\bibinfo {author} {\bibnamefont {{Synopsys
  QuantumATK}}},\ }\href@noop {} {\bibinfo {title} {{QuantumATK version
  V2023.12}}},\ \bibinfo {howpublished}
  {\url{https://www.synopsys.com/manufacturing/quantumatk.html}} (\bibinfo
  {year} {2023})\BibitemShut {NoStop}%
\bibitem [{\citenamefont {Smidstrup}\ \emph {et~al.}(2019)\citenamefont
  {Smidstrup}, \citenamefont {Markussen}, \citenamefont {Vancraeyveld},
  \citenamefont {Wellendorff}, \citenamefont {Schneider}, \citenamefont
  {Gunst}, \citenamefont {Verstichel}, \citenamefont {Stradi}, \citenamefont
  {Khomyakov}, \citenamefont {Vej-Hansen} \emph
  {et~al.}}]{smidstrup2019quantumatk}%
  \BibitemOpen
  \bibfield  {author} {\bibinfo {author} {\bibfnamefont {S.}~\bibnamefont
  {Smidstrup}}, \bibinfo {author} {\bibfnamefont {T.}~\bibnamefont
  {Markussen}}, \bibinfo {author} {\bibfnamefont {P.}~\bibnamefont
  {Vancraeyveld}}, \bibinfo {author} {\bibfnamefont {J.}~\bibnamefont
  {Wellendorff}}, \bibinfo {author} {\bibfnamefont {J.}~\bibnamefont
  {Schneider}}, \bibinfo {author} {\bibfnamefont {T.}~\bibnamefont {Gunst}},
  \bibinfo {author} {\bibfnamefont {B.}~\bibnamefont {Verstichel}}, \bibinfo
  {author} {\bibfnamefont {D.}~\bibnamefont {Stradi}}, \bibinfo {author}
  {\bibfnamefont {P.~A.}\ \bibnamefont {Khomyakov}}, \bibinfo {author}
  {\bibfnamefont {U.~G.}\ \bibnamefont {Vej-Hansen}}, \emph {et~al.},\
  }\href@noop {} {\bibfield  {journal} {\bibinfo  {journal} {Journal of
  Physics: Condensed Matter}\ }\textbf {\bibinfo {volume} {32}},\ \bibinfo
  {pages} {015901} (\bibinfo {year} {2019})}\BibitemShut {NoStop}%
\bibitem [{\citenamefont {Van~Setten}\ \emph {et~al.}(2018)\citenamefont
  {Van~Setten}, \citenamefont {Giantomassi}, \citenamefont {Bousquet},
  \citenamefont {Verstraete}, \citenamefont {Hamann}, \citenamefont {Gonze},\
  and\ \citenamefont {Rignanese}}]{van2018pseudodojo}%
  \BibitemOpen
  \bibfield  {author} {\bibinfo {author} {\bibfnamefont {M.~J.}\ \bibnamefont
  {Van~Setten}}, \bibinfo {author} {\bibfnamefont {M.}~\bibnamefont
  {Giantomassi}}, \bibinfo {author} {\bibfnamefont {E.}~\bibnamefont
  {Bousquet}}, \bibinfo {author} {\bibfnamefont {M.~J.}\ \bibnamefont
  {Verstraete}}, \bibinfo {author} {\bibfnamefont {D.~R.}\ \bibnamefont
  {Hamann}}, \bibinfo {author} {\bibfnamefont {X.}~\bibnamefont {Gonze}},\ and\
  \bibinfo {author} {\bibfnamefont {G.-M.}\ \bibnamefont {Rignanese}},\
  }\href@noop {} {\bibfield  {journal} {\bibinfo  {journal} {Computer Physics
  Communications}\ }\textbf {\bibinfo {volume} {226}},\ \bibinfo {pages} {39}
  (\bibinfo {year} {2018})}\BibitemShut {NoStop}%
\end{thebibliography}%



\end{document}


\title{Supporting information for ``Tailored 1D/2D Van der Waals Heterostructures for Unified Analog and Digital Electronics"}

\setcounter{footnote}{1}
\author{Bipul Karmakar}
\altaffiliation{These authors contributed equally to this work}
\affiliation{School of Physical Sciences, Indian Association for the Cultivation of Science, 2A $\And$ B Raja S. C. Mullick Road, Jadavpur,
Kolkata 700032, India}
\author{Bikash Das}
\altaffiliation{These authors contributed equally to this work}
\affiliation{School of Physical Sciences, Indian Association for the Cultivation of Science, 2A $\And$ B Raja S. C. Mullick Road, Jadavpur,
Kolkata 700032, India}
\author{Shibnath Mandal}
\affiliation{School of Physical Sciences, Indian Association for the Cultivation of Science, 2A $\And$ B Raja S. C. Mullick Road, Jadavpur,
Kolkata 700032, India}
\author{Rahul Paramanik}
\affiliation{School of Physical Sciences, Indian Association for the Cultivation of Science, 2A $\And$ B Raja S. C. Mullick Road, Jadavpur,
Kolkata 700032, India}
\author{Sujan Maity}
\affiliation{School of Physical Sciences, Indian Association for the Cultivation of Science, 2A $\And$ B Raja S. C. Mullick Road, Jadavpur,
Kolkata 700032, India}
\author{Tanima Kundu}
\affiliation{School of Physical Sciences, Indian Association for the Cultivation of Science, 2A $\And$ B Raja S. C. Mullick Road, Jadavpur,
Kolkata 700032, India}
\author{Soumik Das}
\affiliation{School of Physical Sciences, Indian Association for the Cultivation of Science, 2A $\And$ B Raja S. C. Mullick Road, Jadavpur,
Kolkata 700032, India}
\author{Mainak Palit}
\affiliation{School of Physical Sciences, Indian Association for the Cultivation of Science, 2A $\And$ B Raja S. C. Mullick Road, Jadavpur,
Kolkata 700032, India}
\author{Koushik Dey}
\affiliation{School of Physical Sciences, Indian Association for the Cultivation of Science, 2A $\And$ B Raja S. C. Mullick Road, Jadavpur,
Kolkata 700032, India}
\author{Kapildeb Dolui}
\affiliation{Lomare Technolgies Limited, 6 London Street, London EC3R 7LP, United Kingdom}
\affiliation{Department of Materials Science $\And$ Metallurgy, University of Cambridge, 27 Charles Babbage Road, Cambridge CB3 0FS, United Kingdom}
\author{Subhadeep Datta}
\altaffiliation{sspsdd@iacs.res.in}
\affiliation{School of Physical Sciences, Indian Association for the Cultivation of Science, 2A $\And$ B Raja S. C. Mullick Road, Jadavpur,
Kolkata 700032, India}
\maketitle
\begin{figure*}[h!]
    \centerline{\includegraphics[scale=0.6, clip]{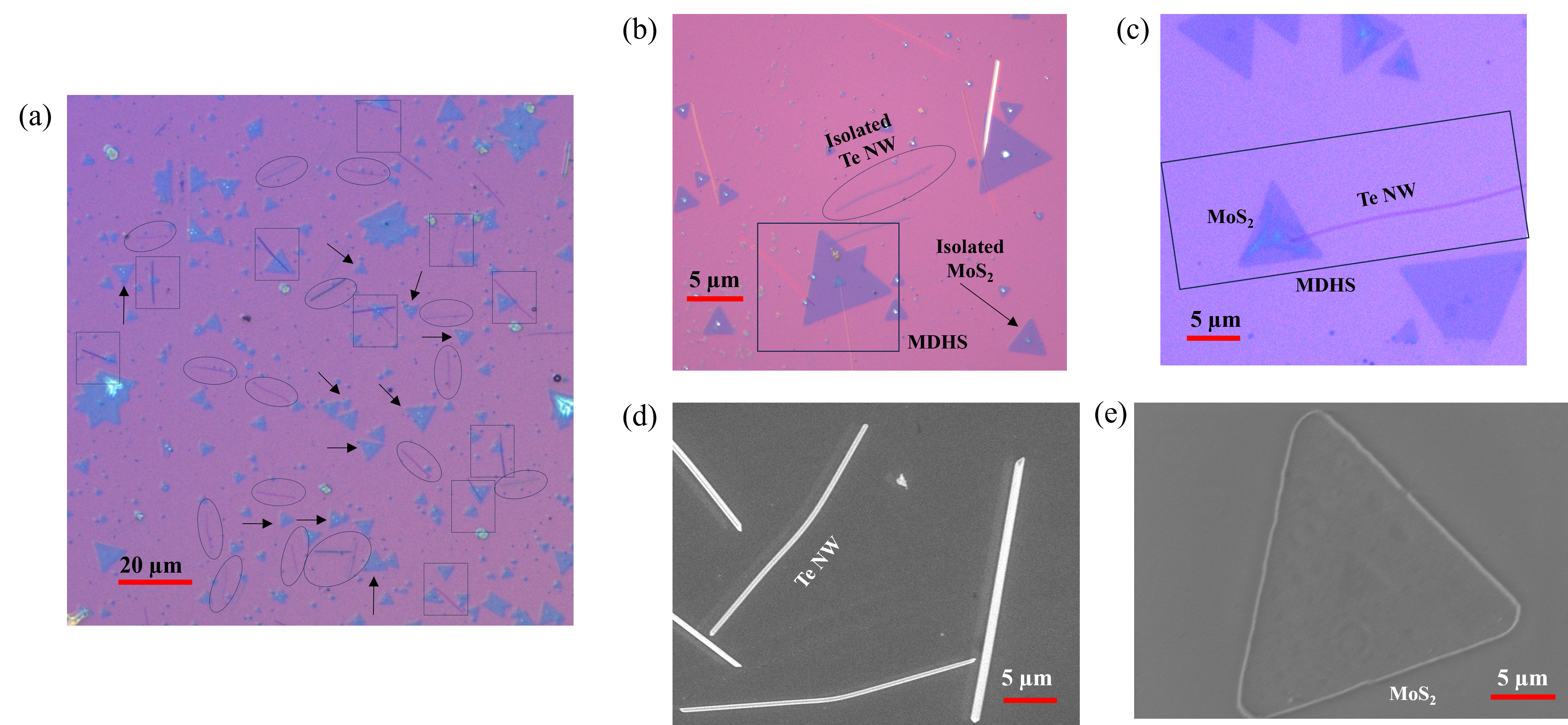}}
    \renewcommand{\thefigure}{S\arabic{figure}}
    \caption{Optical microscope and SEM image. (a), (b) and (c) Optical microscope images of large scale grown microstrucrures. Rectangular boxes, oval shapes and arrows are used to mark the Te NW-MoS$_2$ mixed dimensional heterostructures, isolated Te NWs and isolated MoS$_2$ flakes, respectively. (d) SEM image of Te NWs. (e) SEM image of MoS$_2$ flake.} 
    \label{supp1}
\end{figure*}
\begin{figure*}
    \centerline{\includegraphics[scale=0.6, clip]{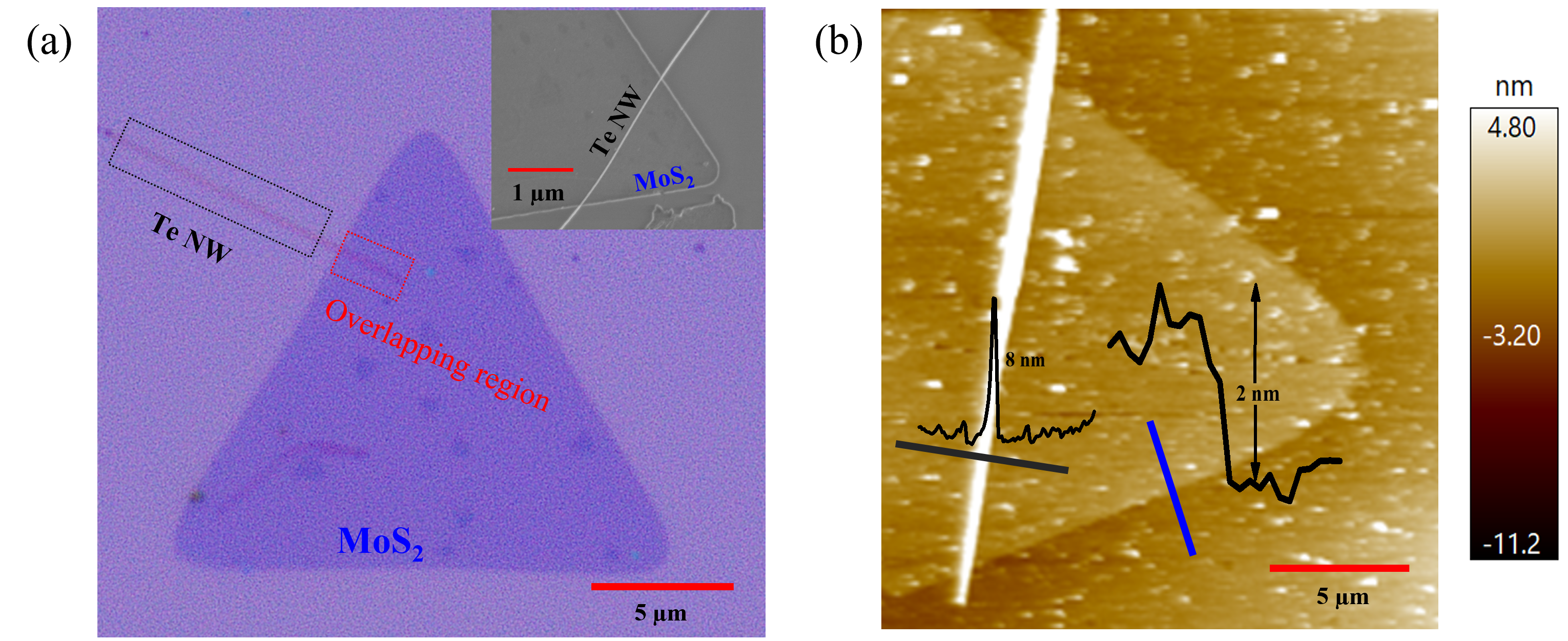}}
    \renewcommand{\thefigure}{S\arabic{figure}}
    \caption{ Morphology study : (a) Optical microscope image of Te NW/MoS$_2$ (1D-2D) heterostructure. Inset: SEM image of a Te wire on a MoS$_2$ flake, confirms a clean heterointerface. (b) AFM image of a 1D-2D heterostructure. The line cuts along the MoS$_2$ and the heterostructure confirm their approximate heights of 2 nm and 8 nm, respectively. } 
    \label{supp2}
\end{figure*}

\begin{figure*}
    \centerline{\includegraphics[scale=0.6, clip]{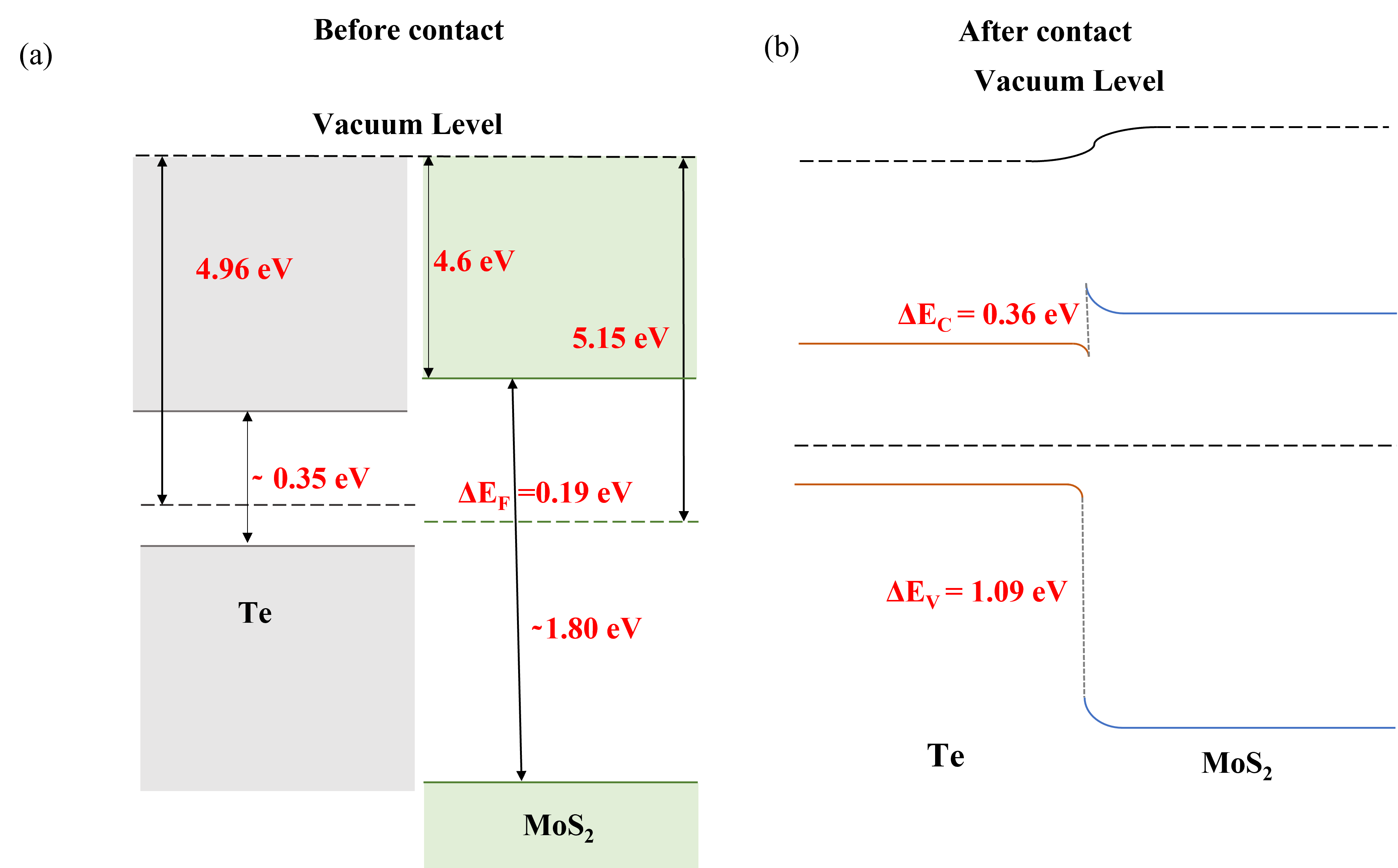}}
    \renewcommand{\thefigure}{S\arabic{figure}}
    \caption{Energy band diagram of Te- MoS$_2$ heterojunction (a) before and (b) after contact, showing a type-I band alignment across the heterojunction. } 
    \label{diode band}
\end{figure*}
\begin{figure*}
    \centerline{\includegraphics[scale=0.6, clip]{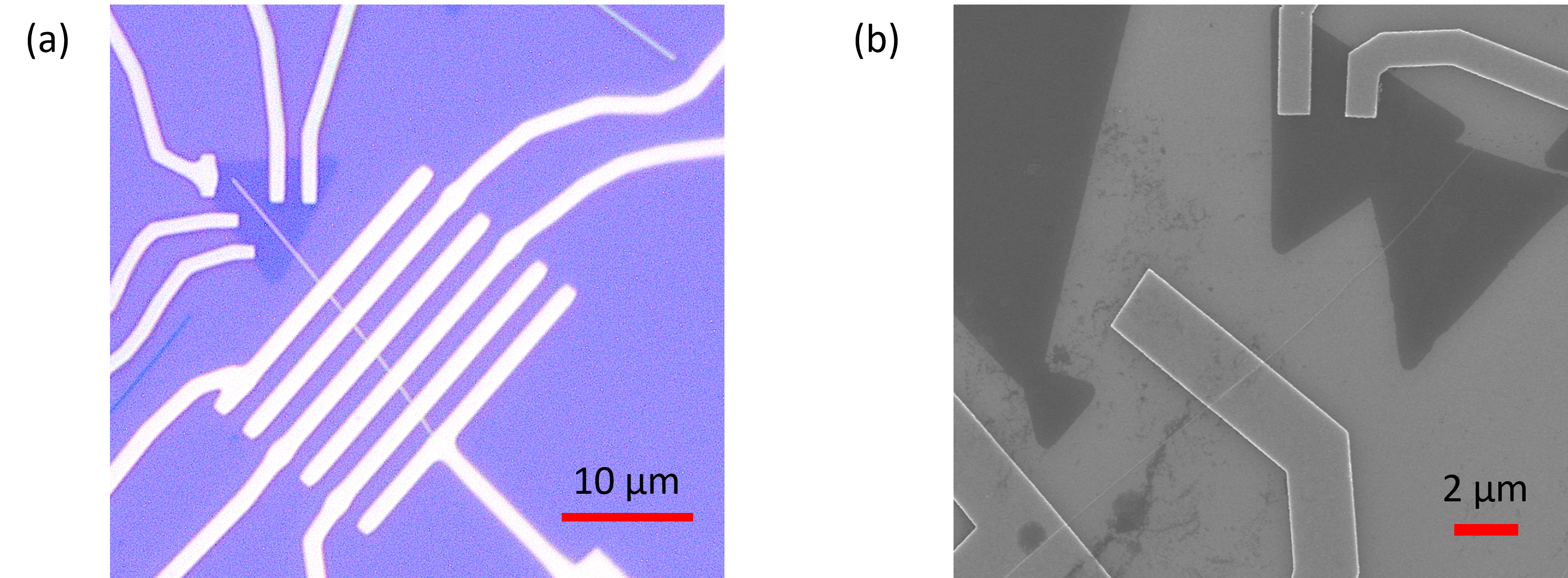}}
    \renewcommand{\thefigure}{S\arabic{figure}}
    \caption{ Representative images of heterostructure devices: (a) Optical microscope image of thick Te NW/MoS$_2$ (1D-2D) heterostructure device (b) SEM image of a thin Te NW/MoS$_2$ heterostructure device } 
    \label{supp12}
\end{figure*}
\begin{figure*}
    \centerline{\includegraphics[scale=0.6, clip]{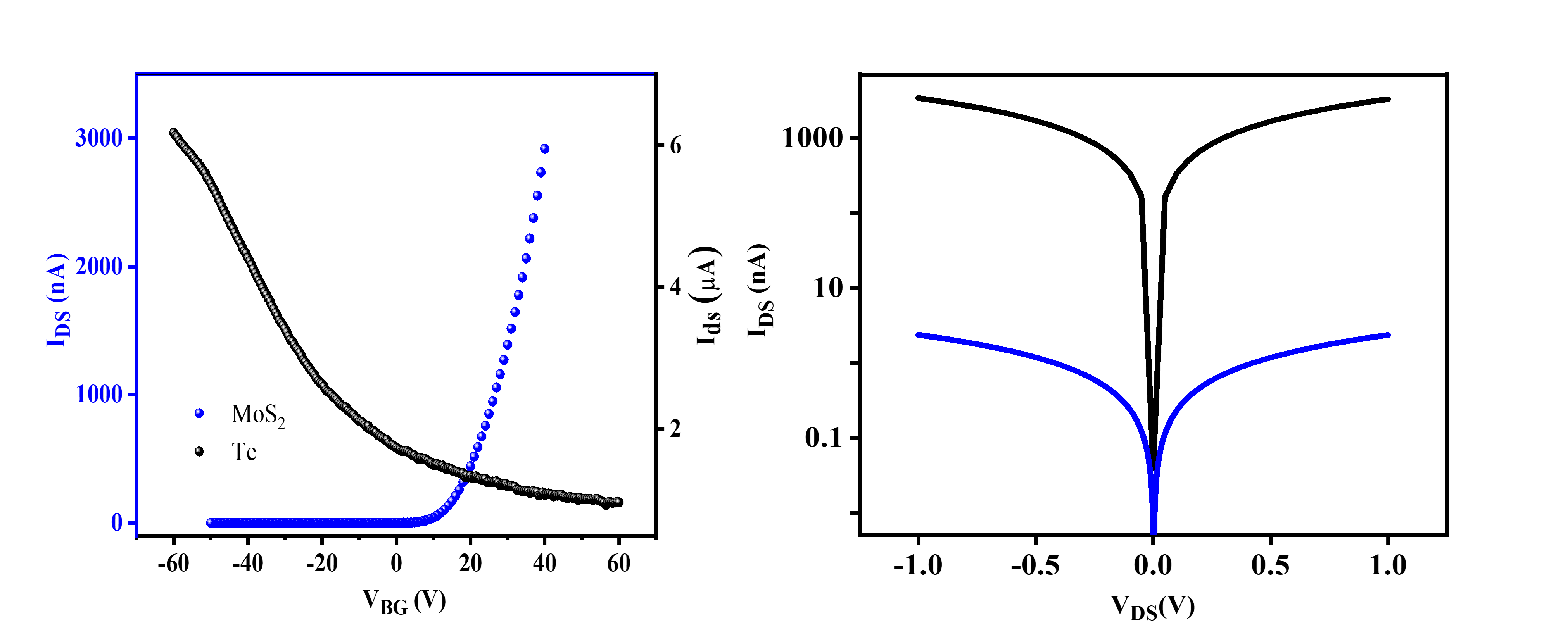}}
    \renewcommand{\thefigure}{S\arabic{figure}}
    \caption{(a) FET transfer characteristics and (b) zero gate I-V charecteristics of individual Te NW (black) and MoS$_2$ (Blue) of the heterostructure shown in Figure 2 (a).} 
    \label{suppiv}
\end{figure*}

\begin{figure*}
    \centerline{\includegraphics[scale=0.8, clip]{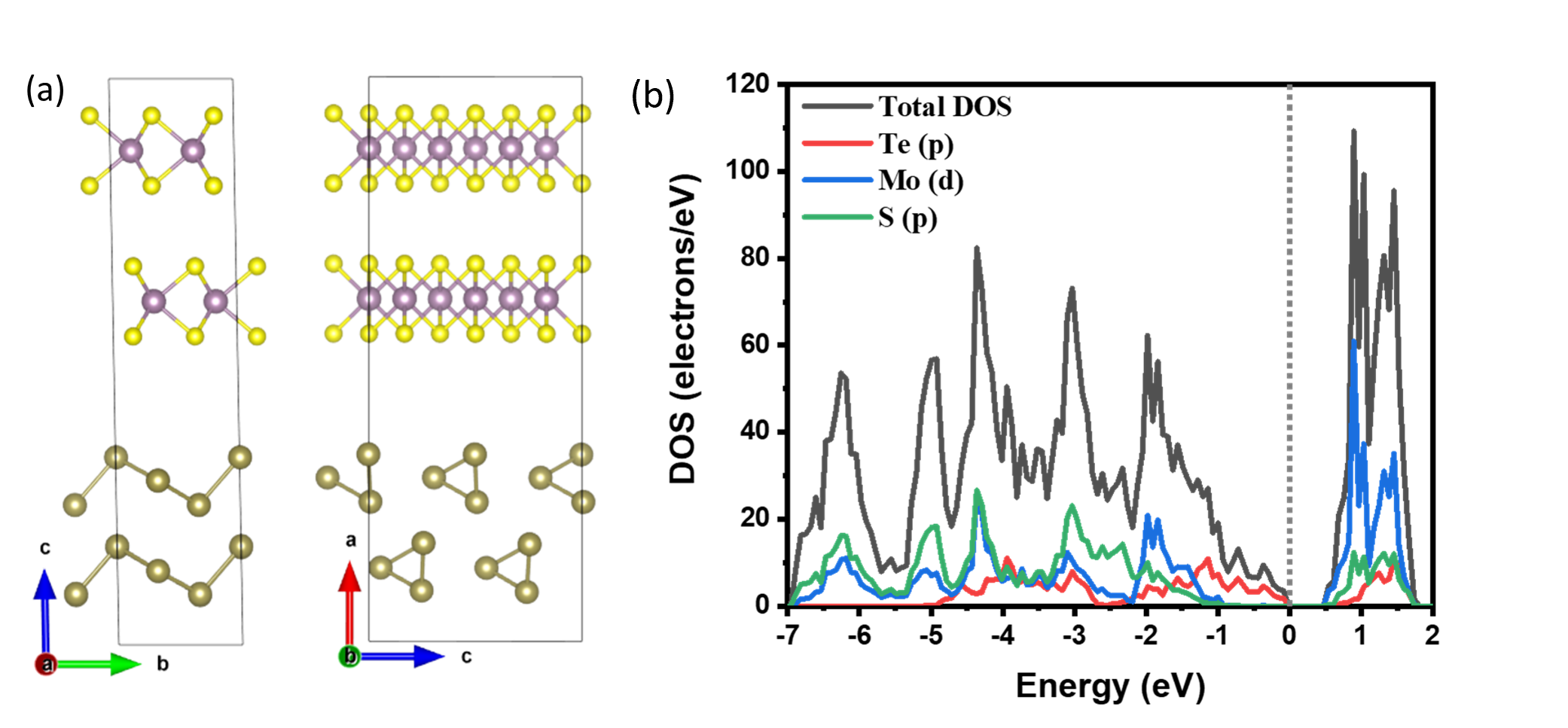}}
    \renewcommand{\thefigure}{S\arabic{figure}}
    \caption{(a) Side view of multilayered Te/MoS2 heterostructure. (b) Orbital-projected density of states representing that valence bands near Fermi level are mostly contributed by Te-5p orbitals and conduction bands near Fermi level are mostly contributed by Mo-4d orbitals (Here the dotted line represents the valence band maximum)} 
    \label{DOS}
\end{figure*}
\begin{figure*}
    \centerline{\includegraphics[scale=0.6, clip]{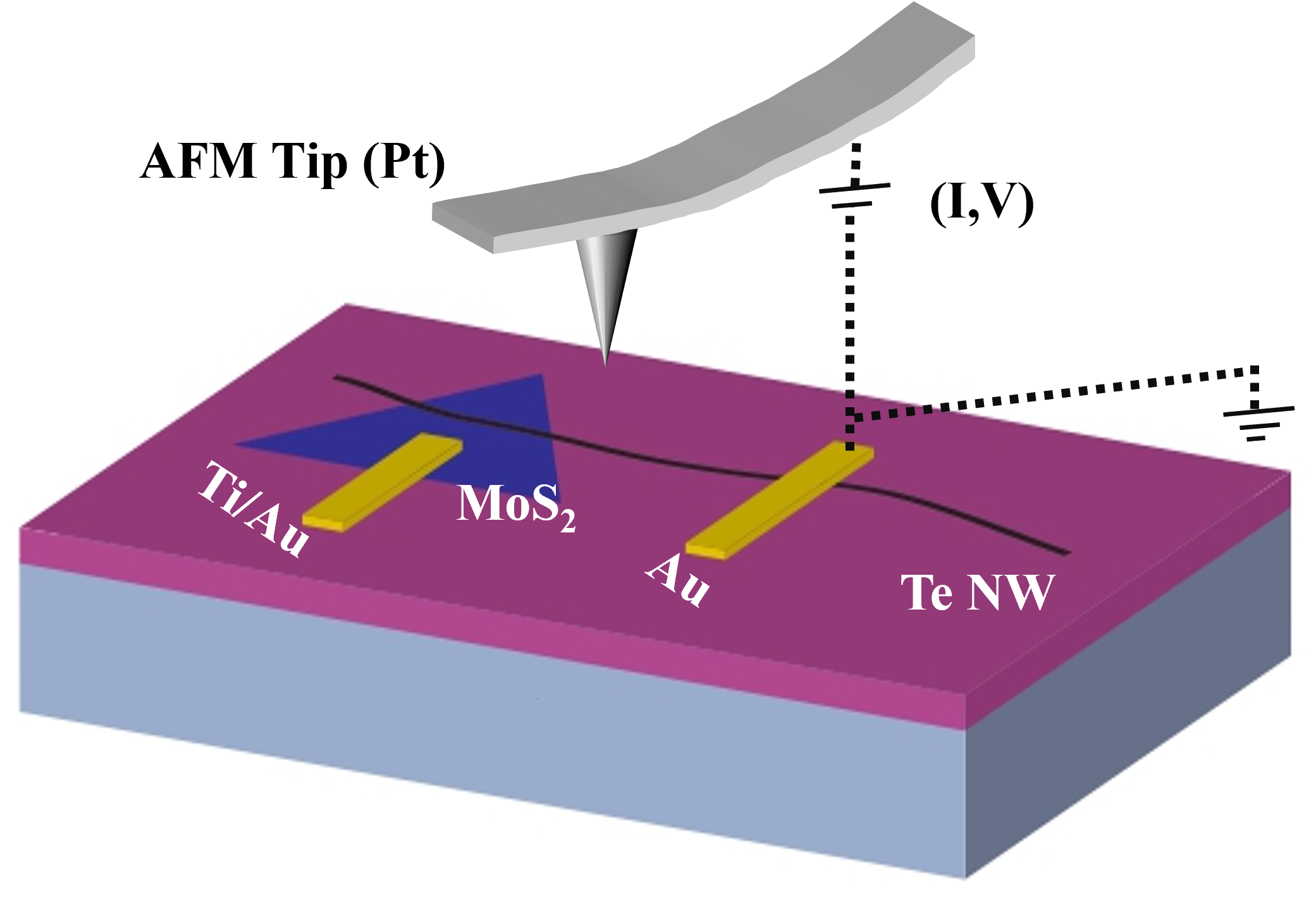}}
    \renewcommand{\thefigure}{S\arabic{figure}}
    \caption{Schematic representation of the CAFM for lateral transport measurement across Te NW of a heterostructure.} 
    \label{Te CAFM}
\end{figure*}

\begin{figure*}
    \centerline{\includegraphics[scale=0.6, clip]{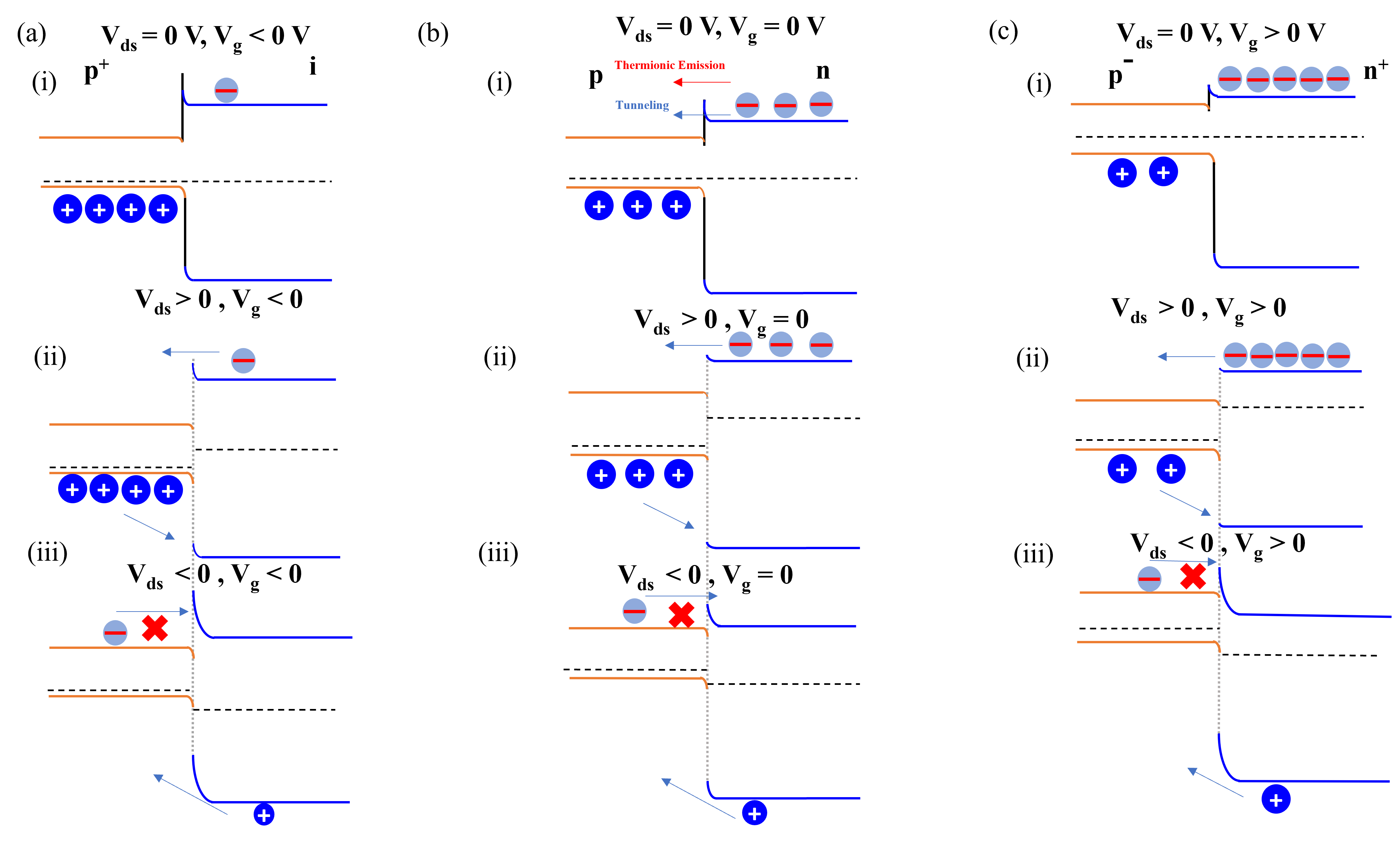}}
    \renewcommand{\thefigure}{S\arabic{figure}}
    \caption{Energy band diagrams of the Te/MoS$_2$ p-n junction diode device. (a), (b) and (c) represents the energy-band diagrams at different drain-source and back-gate voltages.} 
    \label{band diagram}
\end{figure*}